\documentclass[12pt]{iopart}
\usepackage{graphicx}
\usepackage{dcolumn}
\usepackage{bm}
\usepackage{iopams}
\usepackage{color}

\def\be{\begin{equation}}
\def\ee{\end{equation}}
\def\ba{\begin{array}}
\def\ea{\end{array}}
\def\bea{\begin{eqnarray}}
\def\eea{\end{eqnarray}}
\begin{document}
\title[\underline{}]
{Influence of the nuclear matter equation of state on the $r$-mode instability using the finite-range simple effective interaction}
\author{S. P. Pattnaik$^{1,4}$, T. R. Routray$^{1,*}$, X. Vi\~nas$^{2}$, D. N. Basu$^{3}$, M. Centelles$^{2}$, K.Madhuri$^{1,5}$ and B. Behera$^{1}$}

\address{$^1$School of Physics, Sambalpur University, Jyotivihar-768 019, India.}
\address{$^2$Departament de F\'isica Qu\`antica i Astrof\'isica and Institut de Ci\`encies del Cosmos (ICCUB), 
Facultat de F\'isica, Universitat de Barcelona, Mart\'i i Franqu\`es 1, E-08028 Barcelona, Spain}
\address{$^3$Variable Energy Cyclotron Center, 1/AF Bidhan Nagar, Kolkata, 700064, India}
\address{$^4$G.M. Junior College, Sambalpur, Odisha, 768004, India}
\address{$^5$Women's Junior College, Sambalpur, Odisha, 768001, India}
\ead{$^*$ trr1@rediffmail.com (corresponding author)}
\date{\today}

\begin{abstract}
The characteristic physical properties of rotating neutron stars under the $r$-mode oscillation  
are evaluated using the finite-range simple effective interaction. Emphasis is given on examining the influence 
of the stiffness of both the symmetric and asymmetric parts of the nuclear equation of state on these properties. The amplitude 
of the $r$-mode at saturation is calculated using the data of particular neutron stars from the considerations of 
"spin equilibrium" and "thermal equilibrium". The upper limit of the $r$-mode saturation amplitude is found to lie in 
the range $10^{-8}$-$10^{-6}$, in agreement with the predictions of earlier work.
\end{abstract}

\noindent {Manuscript submitted to J. Phys G:Nucl $\&$ Part Phys.}

\vspace{0.5 cm}
\vspace{0.5 cm}


%\submitto{\jpg}
\maketitle

\bigskip
\section{Introduction}
\label{Sec:Int} 
Neutron stars (NSs) may suffer instabilities. Although these instabilities come from different origins, they have  
the general common feature that they can be directly associated with unstable modes of oscillation 
\cite{Andersson1998,Andersson2000,Andersson2001,Andersson2003,Freidman1998,Freidman1999,Lindblom1998,Lindblom2000,
Bildsten2000,Rieutord2000,Ho2011,Alford2012,Mahmoodifar2013,Gusakov2014,Kolomeitsev2014,Mytidis2015,Kokkotas2015,Haskell2015}. 
In the present work the $r$-mode instability is investigated with reference to the nuclear equation of state (EOS). The 
discovery of the $r$-mode oscillation in neutron stars by Andersson \cite{Andersson1998} and confirmed by Friedman and Morsink  
\cite{Freidman1998} opened the window for study of the gravitational waves emitted by NSs by using advanced detecting systems. 
Also it provides a possible explanation for the spin-down mechanism in the hot young NSs as well as for the spin up 
in cold old accreting NSs.

The $r$-mode oscillation results from a perturbation in velocity field of the star. In a non-rotating star, the $r$-modes are 
neutral rotational motions. In a rotating star Coriolis effects provide a weak restoring force that gives them a genuine dynamics. 
The $r$-mode frequency always has different signs in the inertial and rotating frames. That is, although the modes appear to be retrograde 
in the rotating system, an observer in the inertial frame shall view them as prograde. To the leading order, the pattern speed of 
the mode is 
$
\sigma=\frac{\left(l-1\right)\left(l+2\right)}{l\left(l+1\right)} \Omega
$ 
\cite{Papaloizou1978,Andersson1999}. 
Since $0<\sigma<\Omega$ for all modes $l\geq 2$, where $\Omega$ is the angular velocity of the star in the inertial frame, 
the $r$-modes are destabilized by the standard Chandrasekhar-Friedman-Schutz (CFS) mechanism \cite{Chandrasekhar1970,Friedman1978} 
and are unstable because of emission of gravitational waves. The gravitational radiation emitted by the $r$-modes comes from their 
time-dependent mass currents. This is the gravitational analogue of magnetic monopole radiation. The quadrupole $l=2$ $r$-mode is 
more strongly unstable to gravitational radiation than any other mode in neutron stars. Further, these modes exist with velocity 
perturbation if and only if $l=m$ \cite{Andersson1999,Provost1981}. This emission in gravitational waves causes a growth in the 
mode energy $E_{rot}$ in the rotating frame, despite a decrease in the inertial-frame energy $E_{inertial}$. This puzzling effect 
can be understood from the relation between the two energies,
$
E_{rot}=E_{inertial}-\Omega J,
$
where the angular momentum of the star is $J$. From this fact it is clear that $E_{rot}$ may increase even in the case that both 
$E_{inertial}$ and $J$ decrease.

 The instability of the $r$-mode is relevant if it grows faster than it is damped out by the viscosity \cite{Cutler1987}. 
So the time-scale for gravitational driven instability needs to be sufficiently short as compared to the viscous damping time-scale. 
The amplitude of $r$-modes evolves with a time dependence $ e^{i\omega t - t/\tau } $ as a consequence of ordinary hydrodynamics 
and the influence of the various dissipative processes. The real part of the frequency $\omega$ of these modes at the 
lowest order of its expansion in terms of the angular velocity $\Omega$ is given by 
$
\omega=-\frac{\left(l-1\right)\left(l+2\right)}{l+1} \Omega
$
\cite{Lindblom1998, Papaloizou1978}.
 The imaginary part $1/\tau$ is determined by the effects of gravitational radiation, viscosity, etc.  
\cite{Lindblom2000,Lindblom1998,Owen1998}. The time-scales associated with the different processes involve the actual physical 
properties of the neutron star. In computing these time-scales the role of nuclear physics comes into picture, where one gets 
a platform to attempt to constrain the uncertainties existing in the nuclear EOS.

In some studies in this context \cite{Wen2012,Vidana2012,Moustakidis2015}, one of the basic emphases was to examine the influence 
of the slope of nuclear symmetry energy, i.e. the $L$-value, on the $r$-mode instability boundary where the predictions are examined with 
the data of the NS pulsars in Low-mass X-ray binaries (LMXBs) and millisecond radio pulsars (MSRPs). In the present work we make a 
related study but using the finite-range simple effective interaction \cite{trr1998,trr2002}. We examine the $L$-dependence of 
the instability boundary in the context of observational NS data. In order to study the influence of the nuclear matter stiffness, 
the calculation has also been done for two values of the nuclear incompressibility. 

In Section 2, the condition for the $r$-mode instability is outlined along with the various dissipative mechanisms and the 
respective time-scales. The expression for the spin-down rate is given under the consideration of constant temperature. 
In the same Section we give details about the finite range simple effective interaction (SEI) and its parameter
determination. The neutron star equation of state is obtained using SEI and assuming $n+p+e+\mu$ matter in normal phase. 
The crust-core phase transition is discussed using the thermodynamic method. 
In Section 3, we present our results for the $r$-mode instability boundary and compare with  
earlier findings. The upper limit of the $r$-mode amplitude is calculated using the thermal equilibrium condition in NS pulsars 
and their spin-down rate is computed and compared with the observations where data are available as well as with the predictions 
of earlier works. In the last Section 4 we give a brief summary and conclusion.

\section{Dissipative time scales and stability of the $r$-modes} 
\label{Sec:dtss}
 
 In this work we want to study the impact due to the gravitational radiation 
and the dissipative influence of viscosities on the evolution of the $r$-modes. For this purpose we shall consider the effects 
of the radiation on the evolution of the energy mode, which is expressed as the integral of the fluid perturbation 
\cite{Lindblom1998,Lindblom1999}
\begin{equation}
\widetilde{E}=\frac{1}{2}\int{\left[ \rho \delta \vec{v}.\delta \vec{v}^{*}+\left(\frac{\delta p}{\rho}-\delta \Phi \right)
\delta \rho^{*}\right]}d^{3}r,
\label{eq1}
\end{equation}
with $\rho$ being the mass density of the star, $\delta \vec{v}$, $\delta p$, $\delta \Phi$ and $\delta \rho $ are 
perturbations in the velocity, pressure, gravitational potential and density due to oscillation of the mode. The time scales 
for different processes associated with the $r$-mode oscillation are given by \cite{Lindblom1998},
\begin{equation}
\frac{1}{\tau_{i}}=-\frac{1}{2\widetilde{E}}\left(\frac{d\widetilde{E}}{dt}\right)_{i},
\label{eq2}
\end{equation}
where, the index $i$ refers to the various dissipative mechanisms, i.e., gravitational wave emission and viscosity (bulk, shear and 
viscous dissipation at the boundary layer between the crust and the core).

The expressions of the terms involved in the current and mass multipoles are deduced in Ref \cite{Ipser1991,Thorne1980}. In the small 
angular velocity limit, the energy of the mode in equation (\ref{eq1}) can be reduced to 
a one-dimensional integral \cite{Lindblom1998,Vidana2012}
\begin{equation}
\widetilde{E}=\frac{1}{2}\alpha^{2} R^{-2l+2} \Omega^{2} \int^{R}_{0} \rho(r) r^{2l+2} dr, 
\label{eq3}
\end{equation}
where, $R$ is the radius of the NS, $\alpha$ is the dimensionless amplitude mode parameter, $\Omega$ is the angular velocity of the NS 
and $\rho(r)$ is the radial dependence of the NS mass density. 

The rate of increase or decrease in the mode energy $(\frac{d\widetilde{E}}{dt})$ under gravitational and viscous dissipation processes 
have been computed \cite{Ipser1991}. The time-scale $1/\tau$ of the imaginary part of the $r$-mode oscillation can now be expressed as 
the sum of the contributions of all the different dissipative processes and is given by
\begin{equation}
\frac{1}{\tau(\Omega,T)}=\frac{1}{\tau_{GR}(\Omega)}+\frac{1}{\tau_{BV}(\Omega,T)}+\frac{1}{\tau_{SV}(T)}+\frac{1}{\tau_{VE}(\Omega,T)},
\label{eq4}
\end{equation}
where, $1/\tau_{GR}$, $1/\tau_{BV}$,$1/\tau_{SV}$  and $1/\tau_{VE}$ are the contributions from gravitational radiation, bulk and shear 
viscous time-scales in the fluid core and viscous dissipation in the crust-core boundary layer, respectively. Any other possible 
contribution to the dissipation mechanism of the energy of the $r$-mode can be considered in equation (\ref{eq4}), but in the present 
work we have restricted to the four mechanisms mentioned above. The crucial importance of the viscous dissipation in the crust-core 
boundary layer was shown first by Bildsten and Urshomirsky \cite{Bildsten2000}. In a model where the solid crust is not taken into 
consideration, the dissipation contributions come only from the bulk and shear viscosity of the fluid star and it is referred to as 
"minimal model".

The analytical expression for the gravitational radiation time scale is given as \cite{Thorne1980,Owen1998,Lindblom1998},
\begin{equation}
\frac{1}{\tau_{GR}}=\frac{-32 \pi G \Omega^{2l+2}}{c^{2l+3}} \frac{(l-1)^{2l}}{[(2l+1)!!]^2}\left(\frac{l+2}{l+1}\right)^{(2l+2)} 
\int^{R_{x}}_{0}\rho(r)r^{2l+2} dr \hspace{0.5cm}\left( s^{-1}\right), 
\label{eq5}
\end{equation}
where, $G$ and $c$ are the gravitational constant and the velocity of light. The analytical expression for the bulk viscous time-scale is 
obtained in an approximate way, which is valid for stars rotating with slow frequency \cite{Owen1998,Lindblom1998,Vidana2012},
\begin{eqnarray}
\frac{1}{\tau_{BV}}&=&\frac{4\pi R^{2l-2}}{690}\left(\frac{\Omega}{\Omega_0}\right)^{4}\left(\int^{R_{x}}_{0}\rho(r)r^{2l+2} 
dr\right)^{-1}\nonumber \\ 
&&\times \int^{R_{x}}_{0}\xi_{BV}\left(\frac{r}{R}\right)^{6}\left[1+0.86\left(\frac{r}{R}\right)^{2}\right] r^{2} dr \hspace{0.5cm}
\left( s^{-1}\right), 
\label{eq6}
\end{eqnarray}
where, $\xi$ is the bulk viscosity. The shear viscous dissipation time-scale $1/\tau_{SV}$ is obtained in Ref. \cite{Lindblom1998} and reads
\begin{equation}
\frac{1}{\tau_{SV}}=(l-1) (2l+1) \left(\int^{R_{x}}_{0}\rho(r)r^{2l+2} dr\right)^{-1}\int^{R_{x}}_{0}\eta \hspace{0.1cm} r^{2l} dr 
\hspace{0.5cm}\left( s^{-1}\right), 
\label{eq7}
\end{equation}
 where, $\eta$ is the shear viscosity. 
The upper limit $R_x$ of the integrals (\ref{eq5})--(\ref{eq7}) is $R$, the radius of the fluid star, 
if the effect of the crust is not considered and $R_{c}$, the core radius, if the crust is explicitly taken into account.
The time scale $1/\tau_{VE}$ for viscous dissipation at the boundary layer between the crust 
and the core is given by \cite{Lindblom2000,Bildsten2000,Moustakidis2015}.
\begin{equation}
\frac{1}{\tau_{VE}}=\left[\frac{1}{2\Omega} \frac{2^{l+3/2}(l+1)!}{l(2l+1)!!I_{l}}\sqrt{\frac{2\Omega R_{c}^{2} \rho_{c}}{\eta}}
\int^{R_{c}}_{0} \frac{\rho(r)}{\rho_{c}}\left(\frac{r}{R_{c}}\right)^{2l+2} \frac{dr}{R_c}\right]^{-1} \hspace{0.5cm}\left( s^{-1}\right), 
\label{eq8}
\end{equation}
 where $\rho_c$ is the density at the outer edge of the core. $I_{l}$ in equation ({\ref{eq8}}) 
has the value $I_{2}=0.80411$ for $l=2$ \cite{Lindblom2000}. 

The viscous time scale in equation (\ref{eq8}) is obtained by considering the dissipation in the viscous boundary layer between 
the solid crust and the liquid core under the assumption that the crust is rigid and hence static in the rotating frame \cite{Lindblom2000}.
 The motion of the crust due to mechanical coupling to the core effectively increases $\tau_{VE}$ by $(\frac{ \Delta v}{v})^{-2}$, 
where $\frac{\Delta v}{v}$ is the difference in the velocities in the inner edge of the crust and outer edge of the core divided by 
the velocity of the core \cite{Levin2001}.
In the cases of shear viscosity in the bulk and viscous dissipation at the core-crust boundary, the effects of the viscosity 
come from the electron-electron (ee) and neutron-neutron (nn) scattering. The ee-scattering effect dominates in the temperature 
range $T \leq 10^7$ K, whereas in the range $T< 10^9$ K the nn-scattering is dominant. The respective viscosities $\eta^{ee(nn)}$ are given 
by \cite{Lindblom2000}
\begin{equation}
\eta^{ee}=6 \times 10^6 \left(\frac{\rho}{g \hspace{0.1cm} cm^{-3}}\right)^2 \left(\frac{T}{K}\right)^{-2} \hspace{0.5cm} 
\left({g \hspace{0.1cm} cm^{-1} \hspace{0.1cm} s^{-1}}\right), 
\label{eq9}
\end{equation}
\begin{equation}
\eta^{nn}=347 \left(\frac{\rho}{g \hspace{0.1cm} cm^{-3}}\right)^{9/4} \left(\frac{T}{K}\right)^{-2} \hspace{0.5cm} 
\left({g \hspace{0.1cm} cm^{-1} \hspace{0.1cm} s^{-1}}\right). 
\label{eq10}
\end{equation}
%where all the quantities are in CGS unit and T in K. The bulk viscosity becomes dominant in temperature range T $\geq$ $10^{10}$ K. 
The bulk viscosity in equation (\ref{eq6}) should be computed for the modified URCA process, but here we have used the 
approximate expression used in Refs.\cite{Sawyer1989,Cutler1987,Moustakidis2016}, given by  
 \begin{equation}
\xi_{BV} =6 \times 10^{-59} \left(\frac{l+1}{2}\right)^{2} \left(\frac{Hz}{\Omega}\right)^{2} \left(\frac{\rho}{ g \hspace{0.1cm} 
cm^{-3}}\right)^{2} \left(\frac{T}{ K}\right)^{2} \hspace{0.5cm} 
\left({g \hspace{0.1cm} cm^{-1} \hspace{0.1cm} s^{-1}}\right). 
\label{eq11}
\end{equation} 

In the present study, we will examine the influence of the symmetry energy slope parameter $L$ and of the incompressibility $K$ on the instability window 
and on the saturation value of the $r$-mode amplitude of a pulsar neutron star using the finite-range simple effective interaction (SEI) 
\cite{trr1998,trr2002}. We will compare our results with the predictions of earlier related works 
\cite{Vidana2012,Moustakidis2015,Wen2012,Mahmoodifar2013}. The gravitational radiation tends to drive the $r$-mode to the instability, 
while the viscosity suppresses it. The dissipation effects due to viscosity cause the $r$-mode to decay exponentially as $e^{-t/\tau}$ as long as $\tau > 0$ \cite {Lindblom1998}. In order to make out the role of $\Omega$ and T in various time-scales, it is useful to factor them out by defining respective fiducial time-scales. The time-scale $\tau$ given in the equation (\ref {eq4}) can now be expressed as,
 \begin{eqnarray}
\frac{1}{\tau(\Omega,T)}&=&\frac{1}{\widetilde{\tau}_{GR}}\left(\frac{\Omega}{\Omega_0}\right)^{2l+2}
+\frac{1}{\widetilde{\tau}_{SV}}\left(\frac{10^9 K}{T}\right)^{2}
+\frac{1}{\widetilde{\tau}_{BV}}\left(\frac{\Omega}{\Omega_0}\right)^{2}\left(\frac{T}{10^9 K}\right)^{6}
\nonumber \\
&+&\frac{1}{\widetilde{\tau}_{VE}}  \left(\frac{10^8 K}{T}\right) \left(\frac{\Omega}{\Omega_0}\right)^{1/2},
\label{eq12}
\end{eqnarray}
where, $\Omega_0=\sqrt{ \pi G \bar{\rho}}$, with $\bar{\rho}= 3M/4 \pi R^3$ being the mean density of a NS with mass $M$ and radius 
$R$, and ${\widetilde{\tau}_{GR}}$, ${\widetilde{\tau}_{SV}}$, ${\widetilde{\tau}_{BV}}$ and ${\widetilde{\tau}_{VE}} $ are the 
respective fiducial time-scales that can be defined from equations (\ref{eq5})-(\ref{eq8}).
%. 
At small $\Omega$, the gravitational radiation is small (due to the $\Omega^{2l+2}$ dependence) while the viscosity dominates and 
keeps the mode stable. But for large angular velocity $\Omega$, the gravitational radiation dominates and drives the mode  to the 
instability. For a given mode $l$, the critical angular velocity $\Omega_c$ is obtained from the condition,
\begin{equation}
\frac{1}{\tau(\Omega_c,T)}=0,
\label{eq13}
\end{equation} 
where, $1/\tau$ is given in equation (\ref{eq12}). At a given $T$ and mode $l$, the equation for $\Omega_c$ is a polynomial of order 
$l+1$ in $\Omega_c^2$ and thus each mode has its own characteristic $\Omega_c$ value. Since the smallest mode $l=2$ is the most important 
one, the study is made for this $l=2$ mode, where the critical frequency is obtained from the solution of equation (\ref{eq13}).

 As the angular frequency of the NS exceeds the critical value $\Omega_c$, the mode becomes unstable and the star emits gravitational radiation that takes away the angular momentum and energy, and the star spins down to the region of stability.  
Following the work of Owen et al. \cite{Owen1998}, the evolution of the angular velocity as the angular momentum is radiated to infinity 
by the gravitational radiation is given by
 \begin{equation}
\frac{d\Omega}{dt}=\frac{2\Omega}{\tau_{GR}}\frac{\alpha^2Q}{1-\alpha^2Q},
\label{eq14}
\end{equation}
where, $\alpha$ is the dimensionless $r$-mode amplitude and $Q=\frac{3 \widetilde{J}}{2\widetilde{I}}$ with
 \begin{equation}
 \widetilde{J}=\frac{1}{MR^4}\int^{R}_{0}\rho(r)r^{6} dr
\label{eq15}
\end{equation}
and
 \begin{equation}
 \widetilde{I}=\frac{8\pi}{3MR^2}\int^{R}_{0}\rho(r)r^{4} dr.
\label{eq16}
\end{equation}

The $r$-mode amplitude $\alpha$ is treated as a free parameter whose value varies within a wide range $1-10^{-8}$. Under the consideration 
of a thermal steady state, where the heat generated by the viscous effect is the same as that taken out by neutrino emission 
\cite{Bondarescu2009,Moustakidis2015}, the spin-down rate can be derived from equation (\ref{eq14}) to be,
 \begin{equation}
\frac{d\Omega}{dt}=\textsl{C}\left(\Omega^{-6}_{in}-6t\textsl{C}\right)^{-7/6},
\label{eq17}
\end{equation}
where $C$ is given by the expression $
\textsl{C}=\frac{2\alpha^2Q}{\widetilde{\tau}_{GR}\left(1-\alpha^2Q\right)}\frac{1}{\Omega_0^6}
$
and $\Omega_{in}$ is a free parameter whose value corresponds to be the initial angular velocity. The NS
spin shall decrease until it approaches its critical angular velocity $\Omega_c$. The time $t_c$ taken by the NS to evolve 
from its initial value $\Omega_{in}$ to its minimum value $\Omega_{c}$ is given by
 \begin{equation}
t_c=\frac{1}{6\textsl{C}}\left(\Omega_{in}^{-6}-\Omega_{c}^{-6}\right) .
\label{eq18}
\end{equation}
\subsection{Neutron star EOS using SEI} 
\label{Sec:NSsei}
The finite range simple effective interaction (SEI) constructed in 1998 \cite{trr1998} has been widely used in studies 
of nuclear matter \cite{trr2005,trr2007,trr2009,trr2011} at  zero and finite temperature. This interaction, with a Gaussian 
form factor for the finite range part, has also been used to study  the ground-state properties of spherical and 
deformed nuclei \cite{trr2015,trr2016,trr2013} as well as in the dynamical calculation of fission phenomena \cite{trr2016}. Here, in the 
present study of the $r$-mode oscillation in NSs, we  use again the SEI with a Gaussian form factor. 
The SEI used in this calculation is given by \cite{trr2013}
\begin{eqnarray}
v_{eff}(\mathbf{r})&=&t_0 (1+x_0P_{\sigma})\delta(\mathbf{r}) \nonumber \\
&&+\frac{t_3}{6}(1+x_3 P_{\sigma})\left(\frac{\rho({\bf R})}
{1+b\rho({\bf R})}\right)^{\gamma} \delta(\mathbf{r}) \nonumber \\
&& + \left(W+BP_{\sigma}-HP_{\tau}-MP_{\sigma}P_{\tau}\right)f(\mathbf{r}),
\label{eq19}
\end{eqnarray}
where, $\mathbf{r}=\vec{r}_1-\vec{r}_2$ and $\mathbf{R}=(\vec{r}_1+\vec{r}_2)/2$  are the relative and center of mass
coordinates of the two nucleons and $f(\mathbf{r})$ is the functional form  factor of the finite range interaction,
 which can take any conventional Yukawa, Gaussian or exponential form and depends on a single parameter 
$\alpha_G$, which is the range of the interaction. The zero range density dependent part ($t_3$ term) is modified with 
the denominator $(1+b\rho)$ in order to ensure that the EOS of nuclear matter does not become supraluminous at any density.
Therefore the parameter $b$ is determined by the condition
$b \rho_0 \geq \left[ \left (\frac{m c^2}{T_{f_0}/5-e(\rho_0)}\right)^{1/(\gamma+1)}-1 \right]^{-1} $,
 where $mc^2$ and $T_{f_0}=\frac{\hbar^2k_{f_0}^2}{2m}$ 
with $k_{f_0}=\left(\frac{3\pi^2\rho_0}{2}\right)^{(1/3)}$ are the nucleon mass and the Fermi kinetic energy, respectively, 
with $\rho_0$ being the saturation density and $\gamma$ is the exponent of the density dependent term of the interaction \cite {trr1987}. 
The SEI contains altogether eleven parameters, namely, $t_0$, $x_0$, $t_3$, $x_3$, $b$, $\gamma$, $\alpha_G$, $W$, $B$, $H$ and $M$. 
The energy per particle in asymmetric nuclear matter (ANM) for the SEI with a Gaussian form factor, 
$f(r)=e^{-r^{2}/ \alpha_{G}^{2}}$ is given by \cite{trr2013}
\begin{eqnarray}
e(\rho_n,\rho_p)&=&\frac{3\hslash^2}{10m\rho}\left(k_n^2\rho_n+k_p^2\rho_p\right)
+\frac{\varepsilon_{0}^{l}}{2\rho_0 \rho}\left(\rho_n^2+\rho_p^2\right)
+\frac{\varepsilon_{0}^{ul}}{\rho_0 \rho}\rho_n\rho_p \nonumber \\
&&+\frac{1}{\rho}\left[\frac{\varepsilon_{\gamma}^{l}}{2\rho_0^{\gamma+1}}\left(\rho_n^2+\rho_p^2\right)
+\frac{\varepsilon_{\gamma}^{ul}}{\rho_0^{\gamma+1}}\rho_n\rho_p\right]
\left(\frac{\rho({\bf R})}{1+b\rho({\bf R})}\right)^{\gamma} \nonumber \\
&&+\frac{\varepsilon_{ex}^{l}}{2\rho_0 \rho}\rho_n^2\left[\frac{3\Lambda^6}{16k_n^6}
-\frac{9\Lambda^4}{8k_n^4}+\left(\frac{3\Lambda^4}{8k_n^4}-\frac{3\Lambda^6}{16k_n^6}\right)
e^{-4k_n^2/\Lambda^2} \right] \nonumber \\
&&+\frac{\varepsilon_{ex}^{l}}{2\rho_0 \rho}\rho_p^2\left[\frac{3\Lambda^6}{16k_p^6}
-\frac{9\Lambda^4}{8k_p^4}+\left(\frac{3\Lambda^4}{8k_p^4}-\frac{3\Lambda^6}{16k_p^6}\right)
e^{-4k_p^2/\Lambda^2} \right] \nonumber \\
&& +\frac{\varepsilon_{ex}^{l}}{2\rho_0 \rho} \left[\frac{3\Lambda^3}{2k_n^3} \rho_n^2
\int_0^{2k_n/\Lambda}e^{-t^2}dt + \frac{3\Lambda^3}{2k_p^3} \rho_p^2 \int_0^{2k_p/\Lambda}
e^{-t^2}dt\right] \nonumber \\
&&+\frac{\varepsilon_{ex}^{ul}\rho_n}{\rho_0 \rho}\frac{1}{\Lambda^2}\int_0^{k_p}dkk^2
\bigg[\frac{3\Lambda^4}{8kk_n^3}\left\lbrace e^{-\left(\frac{k+k_n}{\Lambda}\right)^2}
-e^{-\left(\frac{k-k_n}{\Lambda}\right)^2}\right\rbrace \nonumber \\
&& +\frac{3\Lambda^3}{ 4k_n^3 \rho}\int_{\left(\frac{k-k_{n}}
{\Lambda}\right)}^{\left(\frac{k+k_n}{\Lambda} \right)}e^{-t^2}dt \bigg],
\label{eq20} 
\end{eqnarray}
where,  $ \Lambda =\frac{2}{\alpha_G}$, and $\rho_n$ and $\rho_p$ are the neutron and proton densities, 
and $k_n$ and $k_p$ represent the neutron and proton Fermi momentum, respectively, 
$k_i=\left(3\pi^2 \rho_{i}\right)^{1/3}$ $i=n,p$. 
The superscripts 'l' and 'ul' denote the strength of the interaction between like and unlike pairs of nucleons. 
The nine parameters required for the complete study of ANM are $\gamma$, $b$, $\alpha_G$, $\varepsilon_{0}^{l}$, 
$\varepsilon_{0}^{ul}$, $\varepsilon_{\gamma}^{l}$, $\varepsilon_{\gamma}^{ul}$, 
$\varepsilon_{ex}^{l}$ and $\varepsilon_{ex}^{ul}$. 
 The later six new parameters appearing here are 
connected to the interaction parameters $W, B, H, M, t_3, t_0, x_0$ and $x_3$ by the relations given in Ref.\cite{trr2013}. 
In symmetric nuclear matter (SNM), $\rho_n=\rho_p=\rho/2$ and in this case the energy per particle becomes
\begin{eqnarray}
e(\rho)&=&\frac{3\hslash^2k_f^2}{10m}
+\frac{(\varepsilon_{0}^{l}+\varepsilon_{0}^{ul})}{4\rho_0}\rho
+\frac{(\varepsilon_{\gamma}^{l}+\varepsilon_{\gamma}^{ul})}{4\rho_0^{\gamma+1}}
\rho\left(\frac{\rho({\bf R})}
{1+b\rho({\bf R})}\right)^{\gamma} \nonumber \\
 &&+\frac{(\varepsilon_{ex}^{l}+\varepsilon_{ex}^{ul})}{4\rho_0}
\rho\bigg[\frac{3\Lambda^6}{16k_f^6}-\frac{9\Lambda^4}{8k_f^4}
+\left(\frac{3\Lambda^4}{8k_f^4}-\frac{3\Lambda^6}{16k_f^6}\right)
e^{-4k_f^2/\Lambda^2} \nonumber \\
&& + \frac{3\Lambda^3}{2k_f^3}\int_0^{2k_f/\Lambda}e^{-t^2}dt \bigg] 
\label{eq21}
\end{eqnarray}
where, $\rho=\rho_n+\rho_p$ is the nuclear matter density, $k_f=\left(\frac{3\pi^2}{2}\rho\right)^{1/3}$ is the Fermi momentum and
\begin{eqnarray}
\left(\frac{\varepsilon_{0}^{l}+\varepsilon_{0}^{ul}}{2}\right)=\varepsilon_0,    
\left(\frac{\varepsilon_{\gamma}^{l}+\varepsilon_{\gamma}^{ul}}{2}\right)=\varepsilon_{\gamma},   
\left(\frac{\varepsilon_{ex}^{l}+\varepsilon_{ex}^{ul}}{2}\right)=\varepsilon_{ex}.
\label{eq22}
\end{eqnarray}
The study of SNM requires only six parameters $\gamma$, $b$, $\alpha_G$, $\varepsilon_{ex}$, 
$\varepsilon_{0}$ and $\varepsilon_{\gamma}$. The parameters  which describe the ANM and SNM are adjusted carefully using appropriate experimental/empirical constraints as we outline in the following. 

The parameters $\varepsilon_{ex}$ and $\alpha_G$, associated with the exchange part of the energy expression, are determined adopting a simultaneous minimization procedure \cite{trr1998} subject to the constraint that the attractive nucleonic mean field changes sign for a kinetic energy of 300 MeV of the incident nucleon \cite {ber1988,gale1990}. 
With the knowledge of these two parameters, $\varepsilon_{ex}$ and $\alpha_G$, one can compute the momentum dependence of the mean field, which compares well with the predictions of the realistic interaction UVI4+UVII \cite{Wiringa1988} over a wide range of momentum and density \cite{trr1998,trr2002}. The two parameters $\varepsilon_{0}$ and $\varepsilon_{\gamma}$ are determined from the saturation conditions, where the standard values $e(\rho_0)$=16 MeV, $T_{f_0}$=37 MeV (corresponding to $\rho_0$=0.161 fm$^{-3}$) are used. 
The parameter $\gamma$ of the density dependent $t_3$-term is kept as a free parameter allowing all values of $\gamma$ for which the pressure-density curve passes through the region extracted from the analysis of high-energy heavy-ion collision data \cite{Damielewicz2002}. 

The study of ANM requires to know the splitting of the strength parameters  $\varepsilon_{ex}$, $\varepsilon_{\gamma}$ and 
$\varepsilon_{0}$ into the like ($l$) and unlike ($ul$) channels. 
The splitting of the exchange strength parameter $\varepsilon_{ex}$ into like and unlike channels is decided from the condition that the entropy per particle in pure neutron matter (PNM) should not exceed that of  symmetric matter SNM \cite{trr2011}. This prescribes a limiting value for the splitting of $\varepsilon_{ex}$ given by $\varepsilon_{ex}^{l}=\frac{2\varepsilon_{ex}}{3}$. 
 With this partition of $\varepsilon_{ex}$, the $n$ and $p$ effective mass splitting in ANM at saturation density as a function of isospin asymmetry $\beta=\frac{\rho_{n}-\rho_{p}}{\rho_{n}+\rho_{p}}$ 
calculated with SEI \cite{trr2013} compares well over the whole range of asymmetries with the microscopic Dirac-Brueckner-Hartree-Fock (DBHF) prediction \cite{Sammarruca2010}. The splitting of the remaining two strength parameters $\varepsilon_{0}$ and $\varepsilon_{\gamma}$ into like and unlike channels is decided by assuming a standard value of the symmetry energy coefficient $E_s(\rho_0)$, and varying  $E_s^{\prime}(\rho_0)=\rho_{0} \frac{dE_{s}(\rho)}{d\rho}|_{\rho=\rho_0}$, subject to the condition that the asymmetric contribution of the nucleonic part in charge neutral $\beta$-equilibrated n+p+e+$\mu$ matter, referred to as neutron star matter (NSM), be maximum over a wide range of density, taken here to be $10 \rho_0$ \cite{trr2007}.
 This asymmetric contribution of the nucleonic part is defined as $S^{NSM}(\rho)=e(\rho,Y_p)-e(\rho,Y_p=1/2)$.
 
 The slope parameter of the symmetry energy is $L(\rho_0)=3E_s^{\prime}(\rho_0)$. The density dependence of nuclear symmetry energy thus predicted is neither very stiff nor soft and also does not allow direct URCA cooling in typical NSs. The parameter $x_0$ is determined by imposing that
the effective mass splitting between spin-up and spin-down neutron in spin polarized neutron matter reproduce the DBHF predictions 
\cite{trr2015}. Finally, the $t_0$ parameter as well as the spin-orbit strength $W_0$, used in finite nuclei calculations, are fitted to
reproduce the binding energy of $^{40}$Ca and $^{208}$Pb.

  The finite nuclei study using SEI \cite {trr2013,trr2015,trr2016} predicts characteristic values for $E_{s}(\rho_0)$ and $\rho_0$, for EOSs 
having different incompressibilities $K(\rho_0)$, for which the energies and radii over the nuclear chart are reproduced with minimum root 
mean square ({\it {rms}}) deviation. The values of $E_s(\rho_0)$ and $T_{f_0}$ vary within the ranges 36-35 MeV and 36.1-35 MeV, respectively 
while $K(\rho_0)$ varies in the range 210-263 MeV. Here for the two considered EOSs the exponents of the 
density-dependent term are $\gamma=$1/2 and 2/3 ($K(\rho_0)$=246 MeV and 263 MeV, respectively). The symmetry energy is chosen 
as $E_s(\rho_0)$=35 MeV and the values of $E_s^{\prime}(\rho_0)$, obtained from the condition of maximum asymmetric contribution of the nucleonic part $S^{NSM}(\rho)$, 
are 25.42 MeV for $L(\rho_0)$=76.26 MeV and 25.77 MeV for $L(\rho_0)$=77.31 MeV.
For the sake of illustration, the symmetry energy $E_s(\rho)$, the equilibrium proton fraction $Y_p=\frac{\rho_p}{\rho}$ and the asymmetric contribution of the nucleonic part $S^{NSM}(\rho)$ are shown as functions of density in the panels (a), (b) and (c) 
of Figure 1, respectively, for the EOS $\gamma$=1/2 with different values of $L(\rho_0)$ in the range 70-110 MeV. The parameters of SEI along with 
the corresponding nuclear matter saturation properties are given in Table 1 for the two EOSs $\gamma$=1/2 and 2/3.

\begin{figure}[ht]
\vspace{1.0 cm}
	\begin{center}
		\includegraphics[width=0.7\columnwidth,angle=0]{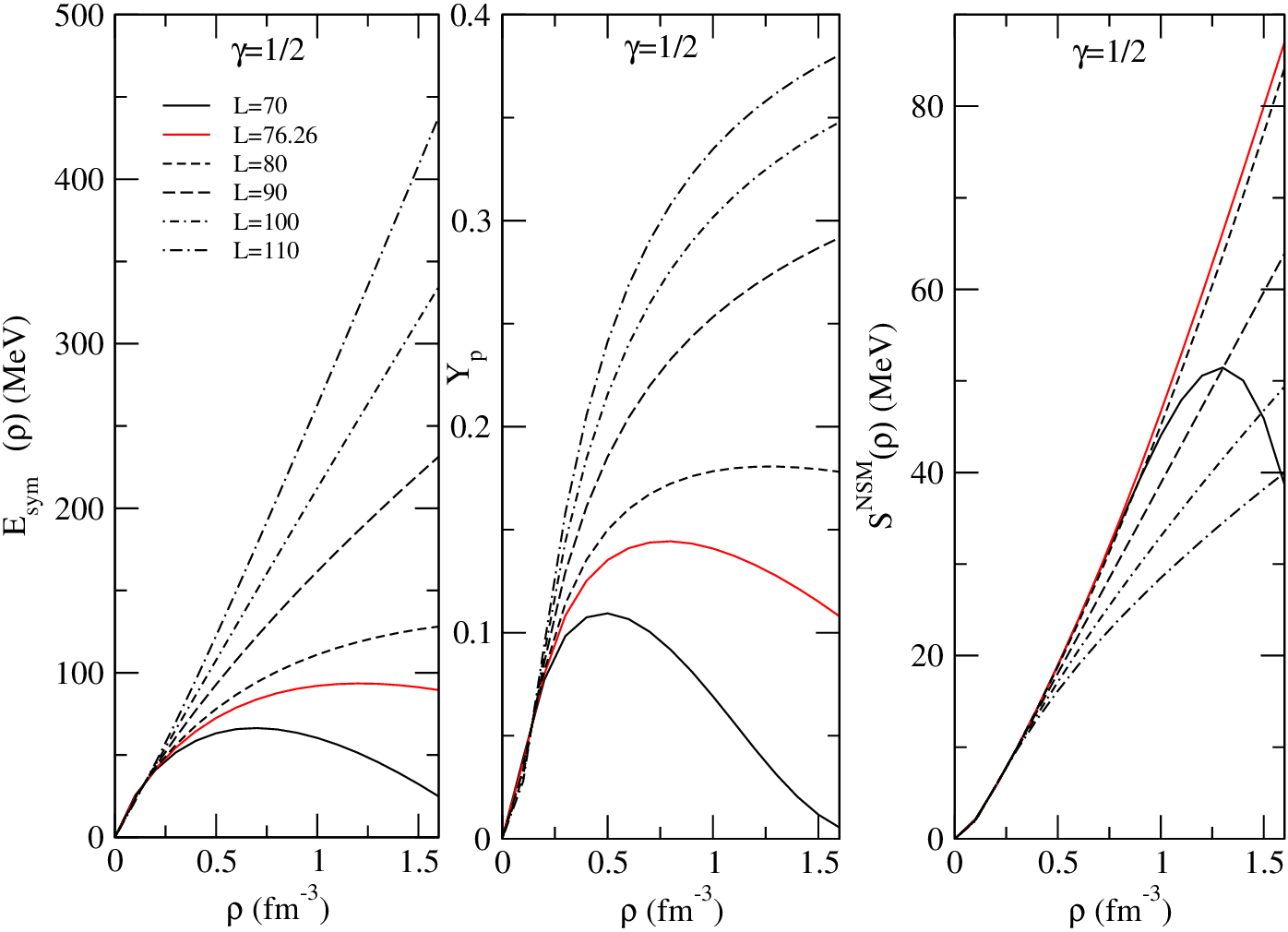}
		\end{center}
	\label{fig:esym}
	\caption{(Colour online) The symmetry energy $E_s(\rho)$ in panel (a), equilibrium proton fraction $Y_p$ in panel (b) 
and asymmetric nucleonic contribution $S^{NSM}(\rho)$ in panel (c) as functions of density $\rho$ for the EOS $\gamma$=1/2 with values 
of slope parameter in the range 70 MeV $ \leq$ $L$ $\leq$ 110 MeV. The curves in red in the three panels correspond to the 
characteristic $E_s^{\prime}(\rho_0)$ value.}
\end{figure}
 \subsection{EOS of neutron star matter}
 \label{Sec:NSMsei}
 The study of the $r$-mode is performed for typical NSs whose core is composed by 
  neutrons, protons, electrons (e) and muons ($\mu$), which are in  $\beta$-equilibrium and fulfill the global charge neutrality condition.
 The equations expressing these conditions are,
 \begin{eqnarray}
\mu_{n}-\mu_{p}=\mu_{e}=\mu_{\mu},
\label{eq23}
  \end{eqnarray}
 and
 \begin{eqnarray}
Y_{p}=Y_{e}+Y_{\mu},
\label{eq24}
  \end{eqnarray}
where, $\mu_{i}$ and $Y_{i}=\frac{\rho_i}{\rho}$, $i=n,p,e,\mu$ are the chemical potentials and particle fraction of neutrons, 
protons, electrons and muons, respectively. The leptonic chemical potentials $\mu_{i},i=e,\mu$ are obtained by considering them in the 
relativistic Fermi gas model, given by
 \begin{eqnarray}
\mu_{e(\mu)}=(c^2\hbar^2k^2_{e(\mu)}+m^2_{e(\mu)}c^4)^{1/2},
\label{eq25}
  \end{eqnarray}
with $k_{e(\mu)}=(3\pi^2\rho_{e(\mu)})^{1/3}=(3\pi^2 \rho Y_{e(\mu)})^{1/3}$ being the electron (muon) Fermi momentum.
The n(p) chemical potentials, given by
$\mu_{n(p)}=\frac{\partial H^{n(p)}(\rho,Y_p)}{\partial \rho_{n(p)}}$ where $H^{n(p)}=\rho_{n(p)}e(\rho,\beta)$, are obtained from the
expression for the energy per particle in ANM given in equation (\ref{eq20}).
The simultaneous solution of equations (\ref{eq23}) and (\ref{eq24}) as a function of density $\rho$ 
predicts the composition of the core.

 The equilibrium proton fraction $Y_p$ thus calculated as a function of density is 
shown in the panel (b) of figure 1 for different values of the slope of the symmetry energy $L$ covering the range $70 MeV \leq L(\rho_0) 
\leq 110 MeV$ for the EOS $\gamma=1/2$ of table 1. 
The energy density $H^{NSM}$ and pressure $P^{NSM}$ of the NSM is now given by 
 \begin{eqnarray}
H^{NSM}=H^{N}(\rho,Y_p)+H^{e}(\rho,Y_p)+H^{\mu}(\rho,Y_p),
\label{eq26}
  \end{eqnarray}
  \begin{eqnarray}
P^{NSM}=P^{N}(\rho,Y_p)+P^{e}(\rho,Y_p)+P^{\mu}(\rho,Y_p),
\label{eq27}
  \end{eqnarray}
where, $H^i$ and $P^i$ for $i=N,e,\mu$ are the nucleonic, electronic and muonic contributions to the energy density and pressure, 
respectively, in the NSM. 
The nucleonic energy density and pressure, $H^{N}$ and $P^N$, are obtained from the expression for the energy per particle in ANM given 
in equation (\ref{eq20}) as   $H^N=\rho e (\rho,Y_p)$ and $P^N=\rho^2 \frac{\partial e (\rho,Y_p)}{\partial \rho}$, for the 
equilibrium value of the proton fraction $Y_p$. The leptonic energy densities and pressure, $H^i$ and $P^i$ for $i=e,\mu$, are obtained 
by treating these systems as a relativistic Fermi gas model. The NS properties are calculated 
using $H^{NSM}$ and $P^{NSM}$ as a function of density in the Tolman-Oppenheimer-Volkov (TOV) equations.
\subsection{Crust-core transition in neutron stars}
 \label{Sec:cctns} 
The crust-core transition in neutron stars is calculated by the thermodynamical method \cite{Kubis2004,Kubis2007}. In this framework, 
the stability condition of the uniform homogeneous core in liquid phase is constructed from the principles of thermodynamics using the 
$\beta$-equilibrated nuclear matter EOS. This has been illustrated in the work of Moustakidis \cite{Moustakidis2010}. The resulting 
stability condition involves the EOS of ANM. Since the isospin dependence in the energy expression is complicated while one works with a finite range effective force, the quadratic approximation of the energy is popularly used that makes the problem relatively ease to handle. However, in an recent work \cite{trr2016} it has been shown explicitly that the quadratic approximation is not 
valid in the low-density, highly-asymmetric regime as the one found in the region of the crust-core transition. In this case it is necessary 
to work using the complete EOS. In order to facilitate the study, the thermodynamical stability condition, expressed in terms of the neutron and proton 
chemical potentials, is given by
\begin{eqnarray}
V_{thermal}&=\frac{\rho}{4}
\Bigg[ 
  \left( 
  \frac{\partial \mu_{n}}{\partial \rho_{n}}
+2\frac{\partial \mu_{n(p)}}{\partial \rho_{p(n)}}
+ \frac{\partial \mu_{p}}{\partial \rho_{p}}
  \right)
+2(1-2Y_{p}) 
 \left(
 \frac{\partial \mu_{n}}{\partial \rho_{n}}
-\frac{\partial \mu_{p}}{\partial \rho_{p}}
 \right) \nonumber\\
&+(1-2Y_{p})^{2}
  \left(
  \frac{\partial \mu_{n}}{\partial \rho_{n}}
-2\frac{\partial \mu_{n(p)}}{\partial \rho_{p(n)}}
+ \frac{\partial \mu_{p}}{\partial \rho_{p}} 
  \right)\nonumber\\
&-\frac{\left\{  (\frac{\partial \mu_{n}}{\partial \rho_{n}} -\frac{\partial \mu_{p}}{\partial \rho_{p}})+(1-2Y_{p})
(\frac{\partial \mu_{n}}{\partial \rho_{n}}-2\frac{\partial \mu_{n(p)}}{\partial \rho_{p(n)}} + \frac{\partial \mu_{p}}
{\partial \rho_{p}})          \right\}^{2}} {\frac{\partial \mu_{n}}{\partial \rho_{n}}-2\frac{\partial \mu_{n(p)}}{\partial \rho_{p(n)}}
+ \frac{\partial \mu_{p}}{\partial \rho_{p}}}
\Bigg]>0 ,
\label{eq28}
 \end{eqnarray}
where, $\mu_{i}=\frac {\partial H(\rho,Y_p)}{\partial \rho_{i}}$ for $i=n, p$.
The matter is uniform and stable so long as $V_{thermal}>0$. The transition density, $\rho_t$, at which this stability condition 
starts being violated, marks the phase transition from uniform homogeneous matter to the inhomogeneous phase indicating the onset 
of nucleonisation and depicts the inner boundary of the neutron star crust.
\begin{table}[ht]
\caption{Values of the nine parameters of ANM for the two EOSs of SEI corresponding to $\gamma=1/2$ and $\gamma=2/3$ together with 
their nuclear matter saturation properties (see text for details).}
\renewcommand{\tabcolsep}{0.05cm}
\renewcommand{\arraystretch}{1.2}
\begin{tabular}{|c|c|c|c|c|c|c|c|c|c|c|c|}\hline
\hline
$\gamma$ & $b$& $\alpha_G$ & $\varepsilon_{ex}$ & $\varepsilon_{ex}^{l}$ & $\varepsilon_{0}$ &
$\varepsilon_{0}^{l}$ & $\varepsilon_{\gamma}$& $\varepsilon_{\gamma}^{l}$ \\
& $\mathrm{fm}$ & $\mathrm{fm}$ & $\mathrm{MeV}$ & $\mathrm{MeV}$ & $\mathrm{MeV}$ & $\mathrm{MeV}$ & $\mathrm{MeV}$ & $\mathrm{MeV}$ \\
\hline
$\frac{1}{2}$& 0.5914 & 0.7597 &-94.4614  &-62.9743 &-78.7832 &-45.8788 &77.5068 &57.76866 \\\hline
$\frac{2}{3}$& 0.78522 &0.7609 &-93.5766  &-62.3844 &-61.9929 &-33.9536 &61.6895 &47.0768 \\\hline
\multicolumn{9}{|c|}{Nuclear matter properties at saturation density} \\
\hline
\multicolumn{1}{|c|}{$\gamma$}&\multicolumn{1}{|c|}{$\rho_0$ ($\mathrm{fm}^{-3}$)} & \multicolumn{2}{c|}{$e (\rho_0) $ (MeV)}
& \multicolumn{1}{c|}{$K (\rho_0)$ (MeV)} & \multicolumn{1}{c|}{$\frac{m^*}{m}(\rho_0,k_{f_0})$}
& \multicolumn{1}{c|}{$E_s (\rho_0)$ (MeV)} & \multicolumn{2}{c|}{$L (\rho_0)$ (MeV)} \\
\hline
\multicolumn{1}{|c|}{$\frac{1}{2}$}& \multicolumn{1}{|c|}{0.1571} & \multicolumn{2}{c|}{-16.0} & \multicolumn{1}{c|}{245.6}
& \multicolumn{1}{c|}{0.7111} & \multicolumn{1}{c|}{35.0} & \multicolumn{2}{c|}{76.26} \\\hline
\multicolumn{1}{|c|}{$\frac{2}{3}$}& \multicolumn{1}{|c|}{0.1552} & \multicolumn{2}{c|}{-16.0} & \multicolumn{1}{c|}{262.6}
& \multicolumn{1}{c|}{0.7119} & \multicolumn{1}{c|}{35.0} & \multicolumn{2}{c|}{77.31} \\\hline
\end{tabular}
\end{table}
\clearpage

\section{Results and Discussion} 
\label{Sec:res}
The NS properties required in the present study of the $r$-mode oscillation are the mass density as a function 
of the distance from the center, the radius of the star, the core-crust transition density and pressure and the radius of the core. These properties 
of the NS are calculated for the two EOSs $\gamma=1/2$ and $2/3$ of table 1, where the realistic crustal EOSs of Feynman, 
Metropolis and Teller \cite{Feynman1949} and Baym, Pethick, and Sutherland \cite{Baym1971} are used for densities below the transition 
density $\rho_t$. As mentioned before, these values of  $\gamma$ correspond to nuclear matter incompressibilities 246 MeV and 263 MeV. 
\begin{figure}[hb]
\vspace{1.0 cm}
	\begin{center}
		\includegraphics[width=0.7\columnwidth,angle=0]{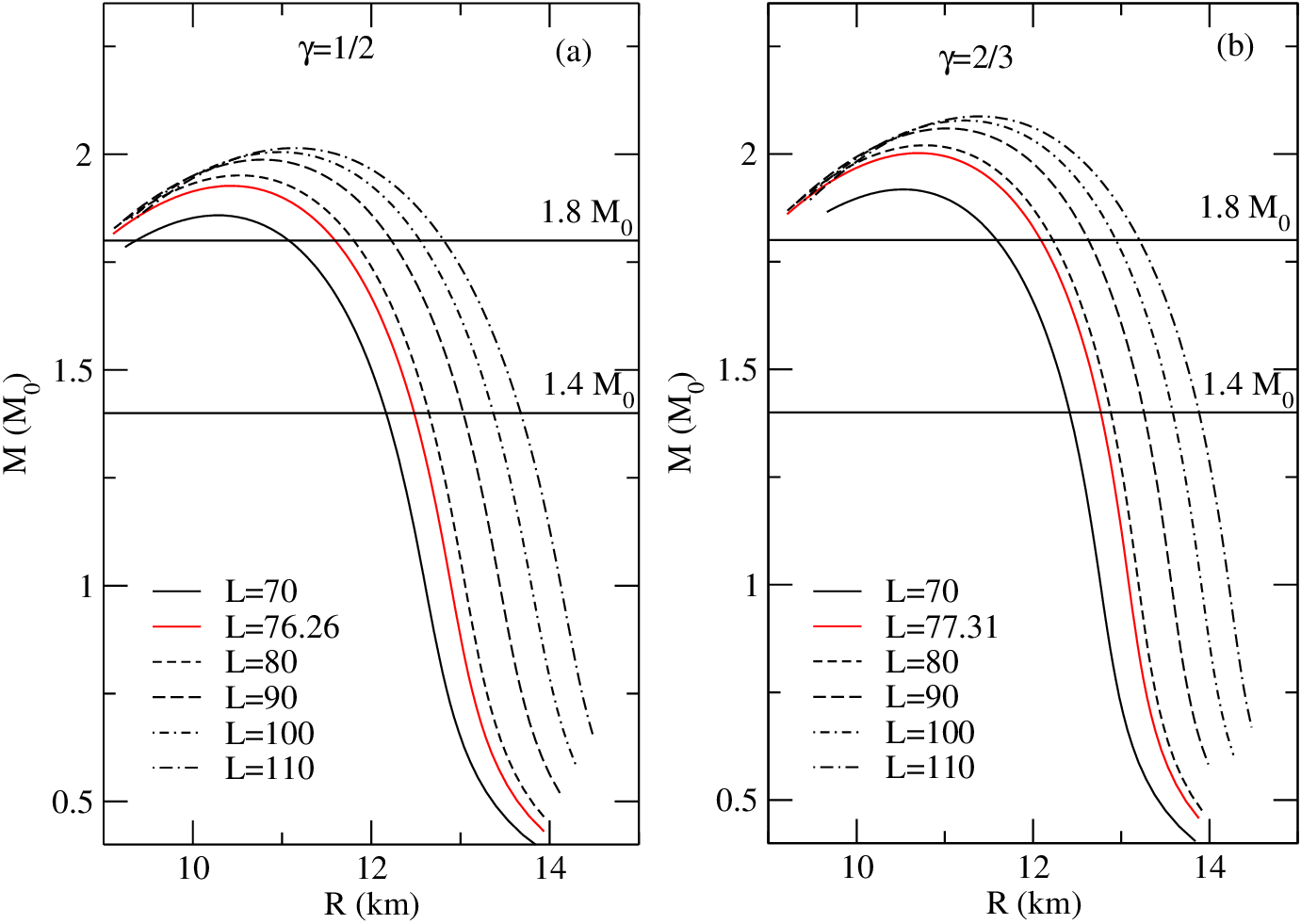}
		\end{center}
	\label{fig:rm}
			\caption{(Colour online) (a) The mass-radius relation for different slope parameters, $L(\rho_0)$, in the range 
70-110 MeV for the EOS $\gamma$=1/2. (b) Same as (a) but for the EOS $\gamma$=2/3.}
\end{figure}
\begin{table} [ht]
\begin{center}
\caption{ Values of the crust-core transition density $\rho_t$ in $fm^{-3}$, pressure $P_t$ in $MeV$ $fm^{-3}$ at the transition density, 
radius $R$ of the NS in km and core radius $R_c$ in km for 1.4 and 1.8 $M_{\odot}$ NSs for the EOSs $\gamma=$1/2 and 2/3. In each case 
three values $L$ are considered in the range 70-110 MeV.}
\renewcommand{\tabcolsep}{0.0991cm}
\renewcommand{\arraystretch}{1.2}
\begin{tabular}{|c|c|c|c|c|c|c|}\hline
L    & $\rho_t$ &  $P_t$    &R& $R_{c} $   &R& $R_{c} $  \\\hline
\multicolumn{1}{|c|}{{(MeV )}}&
\multicolumn{1}{|c|}{{($fm^{-3}$)}}&
\multicolumn{1}{|c|}{{(MeV $fm^{-3}$)}}&
\multicolumn{1}{|c|}{{(km)}}&
\multicolumn{1}{|c|}{{(km)}}&
\multicolumn{1}{|c|}{{(km) }}&
\multicolumn{1}{|c|}{{(km) }}\\

\multicolumn{1}{|c|}{{}}&
\multicolumn{1}{|c|}{{}}&
\multicolumn{1}{|c|}{{}}&
\multicolumn{1}{|c|}{{1.4 $M_{\odot}$}}&
\multicolumn{1}{|c|}{{1.4 $M_{\odot}$}}&
\multicolumn{1}{|c|}{{1.8 $M_{\odot}$}}&
\multicolumn{1}{|c|}{{1.8 $M_{\odot}$}}\\\hline
\multicolumn{7}{|c|}{{$\gamma=1/2$}}\\\hline   
 70.00 & 0.08297  & 0.53228&12.1671 & 11.0649 & 11.0538  &10.4778  	 \\    
 76.26 & 0.07994  & 0.50899&12.4831 & 11.3226 & 11.5834  & 10.9330      \\ 
110.00 & 0.06143  & 0.26836&13.6787 & 12.4269 & 12.7915  & 12.0584      \\\hline
\multicolumn{7}{|c|}{{$\gamma=2/3$}}\\\hline
70.00 & 0.08470     & 0.55109  & 12.4115  & 11.2513   & 11.5648 &10.9078 \\
77.31	&0.08126    &0.52391 & 12.7630  &  11.5394  & 12.0803 & 11.3471 \\
110.00&	0.06450  &0.30082  & 13.8916  &  12.5564  & 13.1910 & 12.3783\\\hline 
\end{tabular}
\end{center}
\end{table}

For each EOS corresponding to a given $\gamma$ value, the NS mass-radius relations are shown in Figure 2 (a) and (b) for different 
values of $L(\rho_0)$ in the range 70-110 MeV.  In these Figures the red line is for the characteristic $L$ value 
provided by the condition of maximum asymmetric contribution of the nucleonic part in NSM, as explained in the fitting protocol 
described in subsection 3.2, while the black lines are obtained by relaxing this condition and imposing a given $L$ value.
 The results of the crust-core transition density $\rho_t$ obtained from the solution of equation (\ref{eq28}), pressure $P_t$ 
at transition density, total radius $R$ and core radius $R_c$  
are given for NSs of masses of 1.4 and 1.8 $M_{\odot}$ in Table 2 for the two EOSs. A linearly decreasing behaviour of $\rho_t$ and $P_t$ 
with increasing value of $L(\rho_0)$ is found for both EoSs, in agreement with the findings of earlier works 
\cite{Moustakidis2015,Wen2012,Gonzalez2017}.
Also it may be seen that for the same $L(\rho_0)$, the EOS having higher value of the incompressibility 
predicts relatively higher values of $\rho_t$ and $P_t$.

The various time-scales in equations (\ref{eq5})-(\ref{eq8}) for the $l=2$ $r$-mode result into the following analytical expressions:
\begin{equation}
\frac{1}{{\tau}_{GR}}=1.3705 \times 10^{-41}\left[I(R_x)_{,6}\right] \left(\frac{\Omega}{Hz}\right)^6 \hspace{0.5cm}\left( s^{-1}\right),
\label{eq29}
\end{equation}
\begin{equation}
\frac{1}{{\tau}^{ee}_{VE}}=\frac{6.9150 \times 10^6}{\left[I(R_x)_{,6}\right]} \left(\frac{R_c}{km}\right)^6 \left(\frac{\rho_{c,14}}{g \hspace{0.1cm} cm^{-3} }\right)^{3/2}\left(\frac{\Omega}{Hz}\right)^{1/2}\left(\frac{K}{T}\right) \hspace{0.5cm}\left( s^{-1}\right),
\label{eq30}
\end{equation}
\begin{equation}
\frac{1}{{\tau}^{nn}_{VE}}=\frac{2.9572 \times 10^6}{\left[I(R_x)_{,6}\right]} \left(\frac{R_c}{km}\right)^6 \left(\frac{\rho_{c,14}}{g \hspace{0.1cm} cm^{-3}}\right)^{13/8}\left(\frac{\Omega}{Hz}\right)^{1/2}\left(\frac{K}{T}\right) \hspace{0.5cm}\left( s^{-1}\right),
\label{eq31}
\end{equation}
\begin{equation}
\frac{1}{{\tau}^{ee}_{SV}}=5.34072069 \times 10^9 \left(\frac{K}{T}\right)^2\frac{\left[I^{ee}(R_x)_{,4}\right]}{\left[I(R_x)_{,6}\right]} \hspace{0.5cm}\left( s^{-1}\right),
\label{eq32}
\end{equation}
\begin{equation}
\frac{1}{{\tau}^{nn}_{SV}}=3.5678\times 10^8 \left(\frac{K}{T}\right)^2\frac{\left[I^{nn}(R_x)_{,4}\right]}{\left[I(R_x)_{,6}\right]} \hspace{0.5cm}\left( s^{-1}\right),
\label{eq33}
\end{equation}
\begin{eqnarray}
\frac{1}{{\tau}_{BV}}&=&4.4177\times 10^{80} \left(\frac{M_{\odot}}{M}\right)^2
\left(\frac{R}{km}\right)^{8}\left(\frac{\Omega}{Hz}\right)^{2}\left(\frac{T}{K}\right)^{6}\nonumber \\ 
&&\times \frac{\left[ \left(\frac{km}{R}\right)^6 I^{BV}(R_x)_{,8}+ \left( \frac{km}{R} \right)^8 I^{BV}(R_x)_{,10}\right]}
{\left[I(R_x)_{,6}\right]} \hspace{0.5cm}\left( s^{-1}\right),
\label{eq34}
\end{eqnarray}
where, $\rho_{c,14}$=$H(\rho_t)/c^2$ in the unit $10^{14} g\hspace{0.1cm} cm^{-3}$. The various $I$-functions appearing in the above equations (\ref{eq29})-(\ref{eq34}) are given by, 
  \begin{eqnarray}
I(R_x)_{,6}=\int_{0}^{R_x} \left[\frac{H(\rho(r))}{MeV fm^{-3}}\right]\left(\frac{r}{km}\right)^6 d\left(\frac{r}{km}\right),
\label{eq35}
  \end{eqnarray}
  \begin{eqnarray}
I(R_x)_{,8}=\int_{0}^{R_x} \left[\frac{H(\rho(r))}{MeV fm^{-3}}\right]\left(\frac{r}{km}\right)^8 d\left(\frac{r}{km}\right),
\label{eq36}
  \end{eqnarray}
  \begin{eqnarray}
I^{nn}(R_x)_{,4}=\int_{0}^{R_x} \left[\frac{H(\rho(r))}{MeV fm^{-3}}\right]^{9/4}\left(\frac{r}{km}\right)^4 d\left(\frac{r}{km}\right),
\label{eq37}
  \end{eqnarray}
  \begin{eqnarray}
I^{ee}(R_x)_{,4}=\int_{0}^{R_x} \left[\frac{H(\rho(r))}{MeV fm^{-3}}\right]^{2}\left(\frac{r}{km}\right)^4 d\left(\frac{r}{km}\right),
\label{eq38}
  \end{eqnarray}
   \begin{eqnarray}
I^{BV}(R_x)_{,8}=\int_{0}^{R_x} \left[\frac{H(\rho(r))}{MeV fm^{-3}}\right]^{2}\left(\frac{r}{km}\right)^8 d\left(\frac{r}{km}\right),
\label{eq39}
  \end{eqnarray}   
   \begin{eqnarray}
I^{BV}(R_x)_{,10}=\int_{0}^{R_x} \left[\frac{H(\rho(r))}{MeV fm^{-3}}\right]^{2}\left(\frac{r}{km}\right)^{10} d\left(\frac{r}{km}\right),
\label{eq40}
  \end{eqnarray}
where, $R_x$ has been defined before (i.e., below Eq.(\ref{eq7})) and $H(\rho(r))$ is the total energy density $H^{NSM}$ in equation (\ref{eq26}) as a function of mass density $\rho(r)$.

The respective fiducial time scales, see Eq.(\ref{eq12}), calculated from equations (\ref{eq29})-(\ref{eq34}) for the crust-core 
model
are given in Table 3 for the EOSs $\gamma$=1/2 and 2/3. It is verified that under the minimal model, the values of ${\widetilde{\tau}_{GR}}$, ${\widetilde{\tau}^{nn(ee)}_{SV}}$ and ${\widetilde{\tau}_{BV}}$ changes maximum within 0.2$\%$ for both the EOSs. 
The temperature dependence of the critical frequency $\nu_c=\Omega_c/2\pi$, 
with $\Omega_c$ being the critical angular velocity  of the rotating NS, is calculated from the solution of equation (\ref{eq13}) in 
conjunction with equation (\ref{eq12}). 
%The critical frequencies are calculated for 1.4 and 1.8 $M_{\odot}$ mass NSs the in case of each EOS 
%$\gamma=1/2$ and $2/3$. For each EOS, three values of $L(\rho_0)$ are considered, $L(\rho_0)=$70 MeV, 110 MeV and the characteristic $L$ 
%value for the EOS given in table 1. The critical frequencies for 1.4 $M_{\odot}$ and 1.8 $M_{\odot}$ mass NSs are shown as a function of 
%the temperature for the EOS $\gamma$=1/2 in the Figures 3 (a) and (b), respectively.
The critical frequencies for NSs of masses 1.4 and 1.8 $M_{\odot}$ are shown as a function of 
the temperature for the EOS $\gamma$=1/2 in Figures 3 (a) and (b), respectively. For each mass, 
three values of $L(\rho_0)$ are considered, namely, $L(\rho_0)=$70 MeV, 110 MeV and the characteristic $L$ value for the EOS $\gamma$=1/2 given in table 1.
The three curves in group (A) in these figures correspond to the case where all the viscous dissipation effects considered in 
equation (\ref{eq4}) are taken into account. The three curves in group (B) of these figures display the results obtained under the "minimal model"
condition, where the viscous dissipation at the crust-core boundary layer is neglected, i.e., $1/\tau_{VE}=0$. The curves of group (A) 
correspond to the perfect rigid crust, i.e., to the case where the crust is co-rotating with the core. But in real situation the inner region of the crust is smeared out due to the presence of pasta phase and there shall be a relative motion between the inner edge of the crust and outer edge of the core. This effect is roughly simulated by introducing the so-called "slippage" factor $S$ where the effective time-scale due to viscous dissipation at crust core boundary layer becomes $\tau_{VE}/S^2$. In principle, the range for possible value of $S$ is 0 to 1. Glampekadis and Andersson \cite{Glampekadis2006} have obtained the realistic value for $S$=0.05. Upon taking this realistic slippage factor, $S$=0.05, into account in equation (\ref{eq13}), the results for the critical frequency $\nu_c$ for the three $L$-values are shown by the curves labelled (C) in Figures 3 (a) and (b). Available information about the spin frequencies of NSs in LMXBs and MSRPs \cite{Mahmoodifar2013,Haskell2012} are also displayed in these figures. Our theoretical predictions about critical frequencies are similar to those reported in Ref. \cite {Haskell2012}, in the sense that all the considered NSs are predicted to lie in the stable region if the hypothesis of rigid crust is assumed.
However, for the minimal model ($1/\tau_{VE}=0$) the instability window is lowered and many of the NSs are predicted to lie in the instability 
region. Considering the elastic property of the viscous boundary layer through the slippage factor $S$, the instability boundary is lowered and 
lies close to the curve predicted by the minimal model if a realistic value of $S=0.05$ is used. The $L$-dependence of the instability boundary
 can be seen from curves in each group (A), (B) and (C) in these figures. In the region of temperature $T<10^9$ K  the critical frequency
is lowered for higher $L$ values, but in the range $T>10^{10}$ K the critical frequency remains almost insensitive to the value of $L$. 
From these considerations we can conclude that the region of instability increases with increasing values of the slope of the symmetry energy, 
$L$.
 \begin{figure}[hb]
\vspace{1.0 cm}
	\begin{center}
		\includegraphics[width=0.9\columnwidth,angle=0,clip=true]{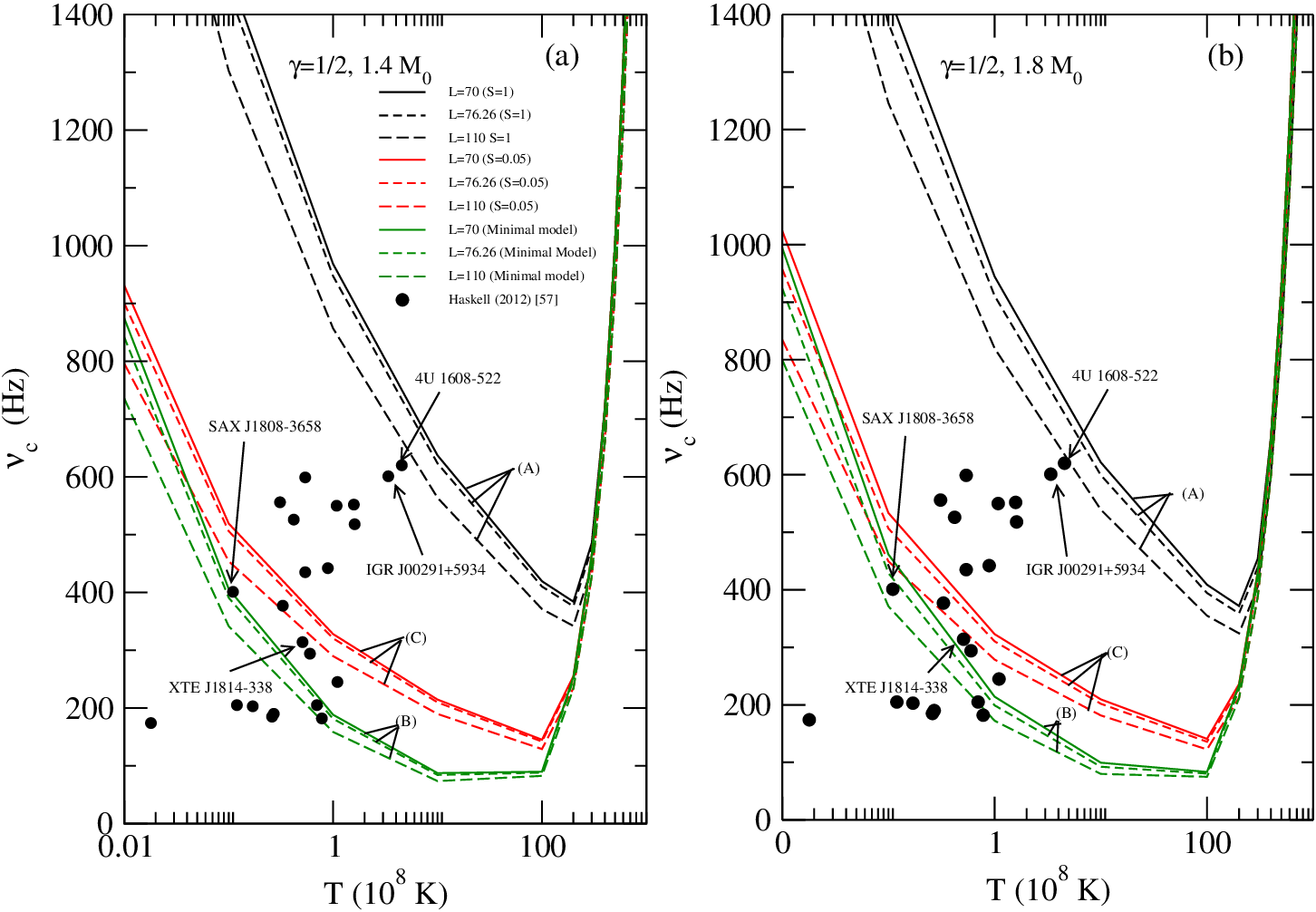}	
		\end{center}
		\caption{(Colour online) (a) Critical frequency $\nu_c$ as a function of temperature $T$ for the EOS $\gamma$=1/2 as $L$ varies from 70 to 110 MeV for 1.4 $M_{\odot}$ NS. 3-curves of (A) are for rigid crust model; 3-curves of (B) are for minimal model; 3-curves of (C) are for penetrating core model accounted for by slippage factor with realistic value $S$ =0.05. (b) Same as (a) but for 1.8 $M_{\odot}$ NS. Legends used for the curves in both panels are the same. }
	\label{fig:graph}
\end{figure}
\begin{table}[ht]
\begin{center}
\caption{The fiducial time scales (in s) for NSs with masses M=1.4 $M_{\odot}$ and 1.8 $M_{\odot}$ for the EOSs $\gamma$=1/2 and 2/3, 
where the results are given, in each case, for 3 values of $L$ in the range 70-110 MeV.}
\renewcommand{\tabcolsep}{0.0991cm}
\renewcommand{\arraystretch}{1.2}
\begin{tabular}{|c|c|c|c|c|c|c|}\hline
$L$&
$\tilde{\tau}_{GR}$&
$\tilde{\tau}_{VE}^{ee}$&
$\tilde{\tau}_{VE}^{nn}$&
$\tilde{\tau}_{BV}$&
$\tilde{\tau}_{SV}^{ee}$&
$\tilde{\tau}_{SV}^{nn}$\\
MeV&s&s&s&s&s&s\\\hline
\multicolumn{7}{|c|}{$\gamma=1/2$ M=1.4 $M_{\odot}$ }\\\hline 
 70.00& -2.6990 & 29.6133 & 66.3949&1.5969$\times10^{11}$  &2.0969$\times10^{8}$ & 5.7591$\times10^{7}$ \\\hline
 76.26& -3.0752 & 30.7745 & 69.3273&1.6772$\times10^{11}$  &2.4203$\times10^{8}$ & 6.5231$\times 10^{7}$   \\\hline
110.00& -4.9424 & 39.9715 & 93.1179&2.0248$\times10^{11}$ & 4.0418$\times10^{8}$ & 1.0191$\times10^{8}$ \\\hline
\multicolumn{7}{|c|}{$\gamma=1/2$ M=1.8 $M_{\odot}$ }\\\hline 
 70.00& -0.5885   &  32.6563      &  73.2175&1.4054$\times10^{11}$ &  9.6814   $\times10^{7}$ &  3.0191$\times10^{7}$\\\hline
 76.26& -0.7358   &  33.8108      &  76.1674&1.5131$\times10^{11}$ &  1.2692 $\times10^{8}$ & 3.8146$\times10^{7}$\\\hline
110.00& -1.2318   &  44.1466      & 102.8443&1.8597$\times10^{11}$ &  2.1836 $\times10^{8}$ & 6.1220$\times10^{7}$\\\hline
\multicolumn{7}{|c|}{$\gamma=2/3$ M=1.4 $M_{\odot}$ }\\\hline 
 70.00& -2.9401 & 28.9290 & 64.6905&1.6328$\times10^{11}$  &2.3608$\times10^{8}$ & 6.3798$\times10^{7}$ \\\hline
 77.31& -3.3814 & 30.2415 & 67.9852&1.7192$\times10^{11}$  &2.7575$\times10^{8}$ & 7.3019$\times 10^{7}$   \\\hline
110.00& -5.3024 & 37.7902 &  87.4939&2.0584$\times10^{11}$ & 4.4760$\times10^{8}$ &1.1115$\times10^{8}$ \\\hline
\multicolumn{7}{|c|}{$\gamma=2/3$ M=1.8 $M_{\odot}$ }\\\hline 
70.00& -0.7203  &  31.5553   &  70.5635  &1.4916$\times10^{11}$ &  1.2721  $\times10^{8}$ &  3.8197$\times10^{7}$\\\hline
 77.31& -0.8908  &   32.8252   &  73.7935  &1.6042$\times10^{11}$ &  1.6348 $\times10^{8}$ &  4.7452$\times10^{7}$\\\hline
110.00& --1.4081 &  41.3611    & 95.7615   &1.9305$\times10^{11}$ & 2.6512 $\times10^{8}$ & 7.2320$\times10^{7}$\\\hline
\end{tabular}
\end{center}
\end{table}

 In order to make it more explicit, the different time-scales, $\tau_{GR}$, $\tau_{SV}$, $\tau_{BV}$ and $\tau_{VE}$ are shown as a 
function of temperature for a given frequency, taken to be $\nu=$ 600 Hz, for NSs of masses 1.4 $M_{\odot}$ and 1.8 $M_{\odot}$ in Figure 4(a) 
and (b) for two values of $L$= 70 and 110 MeV of the EOS $\gamma=1/2$. The gravitational radiation being independent of temperature 
is given by the horizontal line. It can be seen from the curve of viscous dissipation at the crust-core boundary layer $\tau_{VE}$ that 
it is the dominating one in the temperature range T$\leq$ $10^9$ K and effectively prevents the gravitational radiation to render the 
$r$-mode unstable. The bulk viscosity of the fluid core is dominating at higher temperature $T \geq 10^9$ K. Thus the area enclosed 
within the triangle obtained from the intersection of $\tau_{GR}$, $\tau_{VE}$ and $\tau_{BV}$ is the region of instability for the 
given frequency of the star in the rigid crust-core model, ({slippage factor} $S$=1). As we soften the crust-core contribution by 
decreasing $S$ from 1, the curve of  $\tau_{VE}$ moves upward, thereby increasing the area of the triangle. In the minimal model, where 
the crust is neglected, the effect of gravitational radiation is countered by shear viscosity of the fluid star in the range 
T$\leq$ $10^9$ K. This effectively counters the effect of the gravitational radiation below temperature $T \leq 10^7$ K. Therefore, 
in the minimal model, the region of instability represented by the area enclosed in the triangle formed from the intersection of 
$\tau_{GR}$, $\tau_{SV}$ and $\tau_{BV}$ is maximum. Now, on comparing the $L$-dependence of these time-scales from the curves for $L$=70 
and 110 MeV in figures 4(a) and (b), it can be seen that for higher $L$-values the $\tau_{GR}$ takes a relatively small value and moves 
down. The curves for $\tau_{VE}$ and $\tau_{SV}$ move upward, whereas, the bulk viscous time-scale $\tau_{BV}$ remains almost insensitive. 
Due to this behaviour of the time-scales, the area enclosed in the triangles increases with an increase in the $L$ value. This illustrates 
the features relating to the $L$-dependence of the instability window shown in figures 3(a) and (b) for different values of the 
slippage factor $S$. Comparing the results in figures 4(a) and (b), it can be seen that with increasing mass of the NS
 the $L$-dependence of $\tau_{GR}$ and $\tau_{SV}$ becomes relatively more prominent, whereas the influence on the time-scales $\tau_{BV}$ and 
$\tau_{VE}$ is not significant. Due to this fact, the instability window in the temperature range $T\leq$ $10^9$ K lowers by a small extent in the 
case of stars of 1.8 $M_{\odot}$ compared to stars of 1.4 $M_{\odot}$. The pulsar 4U 1608-522 in figure 3(a), which is below the instability 
window for $L$=110 MeV of the rigid crust-core case and predicted to be stable for mass 1.4 $M_{\odot}$, coalesces with the instability window 
for a mass 1.8 $M_{\odot}$ as may be seen from figure 3(b). All these features associated with the dependence of the instability window on the slope 
of the symmetry energy and the NS mass can also be understood from equation (\ref{eq12}) together with the values of the various fiducial 
time-scales reported in table 3. From the values of $\tilde{\tau}_{VE}$, $\tilde{\tau}_{SV}$ and $\tilde{\tau}_{BV}$ for a given value of 
$L$ in any one of the EOS in table 3 it is evident that $1$/$\tilde{\tau}_{VE}$ {(= $1$/$\tilde{\tau}^{nn}_{VE}$ +$1$/$\tilde{\tau}^{ee}_{VE}$) }
is the dominant term in equation (\ref{eq12}) to counter the gravitational radiation effect in the range $T\leq$ $10^9$ K. In absence of 
$\tilde{\tau}_{VE}$, i.e., in the minimal model, the $1$/$\tilde{\tau}_{SV}${(=  $1$/$\tilde{\tau}^{nn}_{SV}$ +$1$/$\tilde{\tau}^{ee}_{SV}$)} 
term counters the effect of gravitational radiation in this range of temperature. The bulk viscosity term $1$/$\tilde{\tau}_{BV}$ takes up 
overshadowing the effects of other viscous terms as $T$ increases beyond $10^9$ K. As $1$/$\tilde{\tau}_{VE}$ $>>$ $1$/$\tilde{\tau}_{SV}$, 
the instability window is raised by a proportionately large extent in the rigid crust model as compared to the minimal model in the range $T\leq$ 
$10^9$ K. Further, since both $1$/$\tilde{\tau}_{VE}$ and $1$/$\tilde{\tau}_{SV}$ decrease with increase in $L$, the instability window in 
both the models is lowered in the range T$\leq$ $10^9$ K. But the increase in $\tilde{\tau}_{BV}$ is marginal as $L$ increases from 70 to 
110 MeV and the instability window practically remains insensitive to the slope of the symmetry energy $l$ in the range $T\geq 10^{10}$ K. A similar behaviour is found using EOSs 
with different incompressibilities, i.e., different values of the $\gamma$ parameter. Now, an increase in the NS mass for a given EOS and given 
$L$ value results into an increase in $\tilde{\tau}_{VE}$ but a decrease in $\tilde{\tau}_{SV}$. Therefore, 
the instability window is lowered in the rigid crust model with increasing mass of the NS, whereas, it is raised in the minimal model.
 All these features are shown in Figures 5(a) and (b) for the EOSs $\gamma$=1/2 and 2/3, respectively, where the instability windows for 
1.4 $M_{\odot}$ and 1.8 $M_{\odot}$ stars are compared for $L$ values of 70 and 110 MeV in both the rigid crust and the minimal model. In order to see 
the influence of $\gamma$, the instability windows for $\gamma$=1/2 and 2/3 are shown in both the models with $L$=70 and 110 MeV in 
Figures 6(a) and (b) for 1.4 $M_{\odot}$ and 1.8 $M_{\odot}$ stars, respectively. It can be realized that with an increase in $\gamma$ the instability 
window follows a lowering trend and this effect becomes more important when the mass of the NS increases.
\begin{figure}[ht]
\vspace{1.0 cm}
\begin{center}
		\includegraphics [width=0.9\columnwidth,angle=0,clip=true]{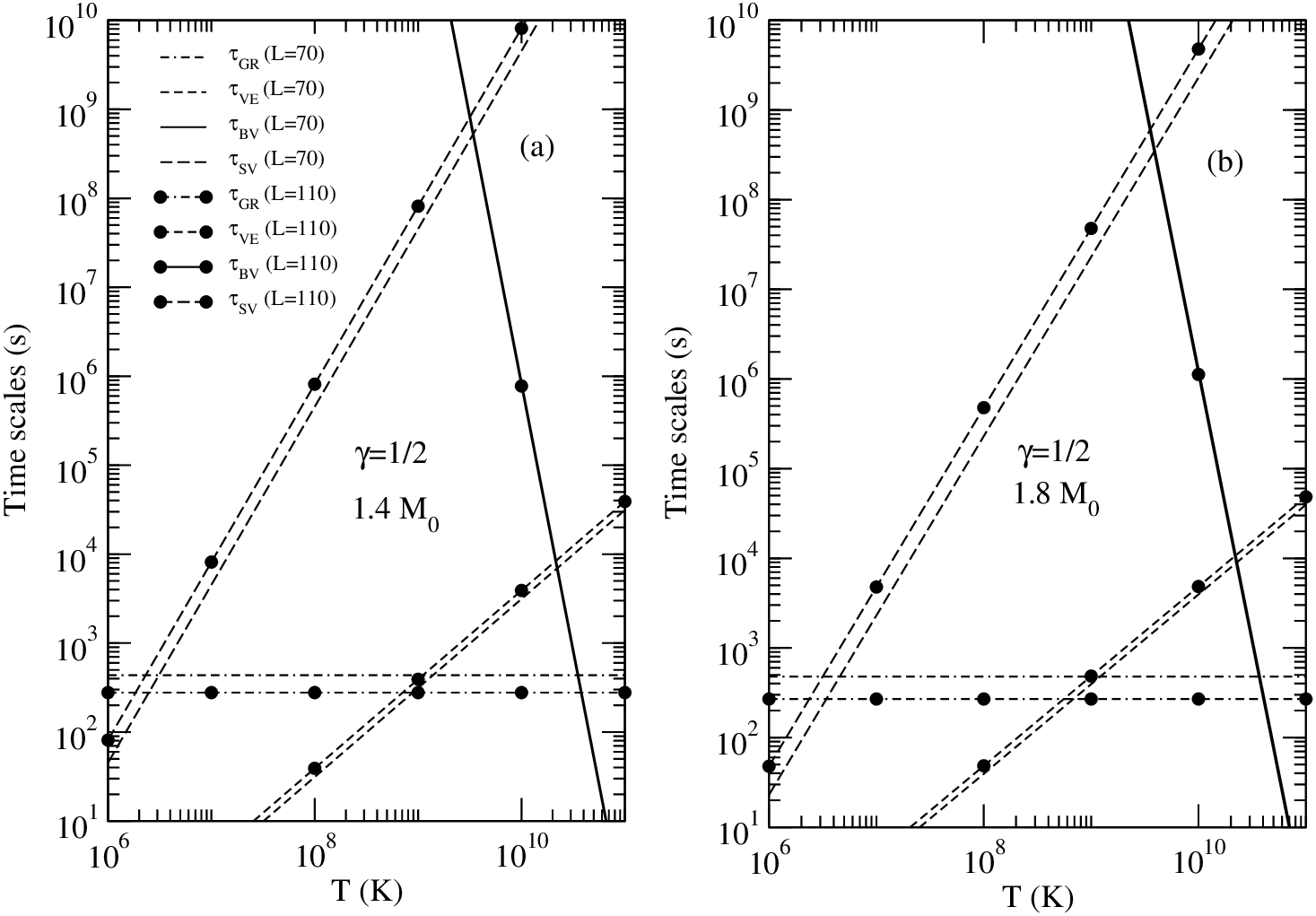}
					\end{center}
					\caption {(a) Different time-scales, $\tau_{GR}$, $\tau_{SV}$, $\tau_{BV}$ and $\tau_{VE}$ as a 
function of temperature $T$ for 1.4 $M_{\odot}$ NS with the EOS $\gamma=$1/2 and two values of $L$=70 and 110 MeV (b) Same as (a) but for 1.8 $M_{\odot}$ NS. Legends used for the curves in both panels are the same. }
	\label{fig:1.4_1.8_12tau}
\end{figure}
\begin{figure}[ht]
\vspace{1.0 cm}
\begin{center}
		\includegraphics [width=0.9\columnwidth,angle=0,clip=true]{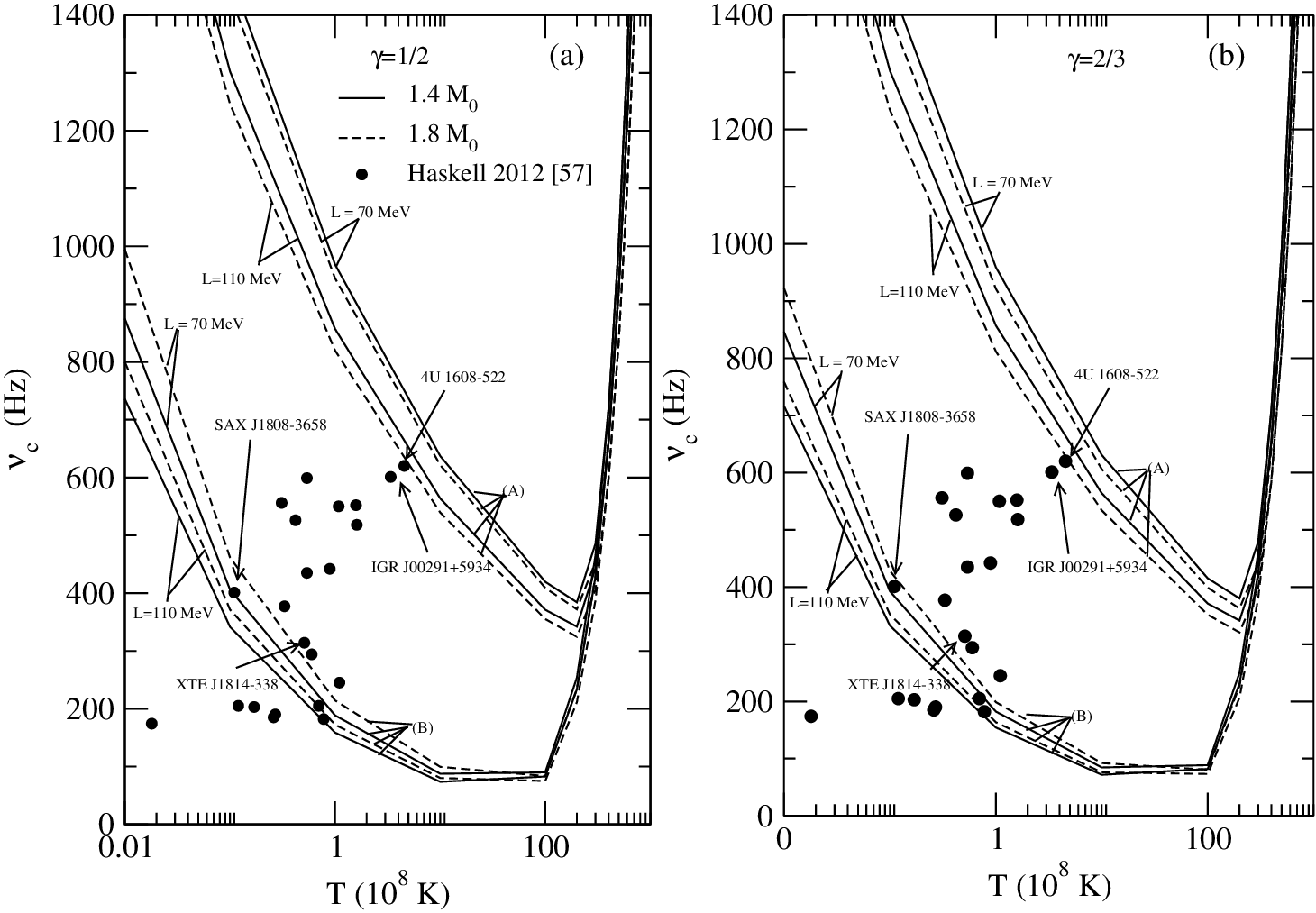}
					\end{center}
					\caption{ (a) Influence of mass of NS pulsar on the instability window for the EOS $\gamma$=1/2. (b) Same as (a) but for the EOS $\gamma$=2/3. Legends for the curves used in both panels are the same. For details, see the text. }
	\label{fig:graph_a1.4_1.8}
\end{figure}
 \begin{figure}[ht]
\vspace{1.0 cm}
\begin{center}
		\includegraphics [width=0.9\columnwidth,angle=0,clip=true]{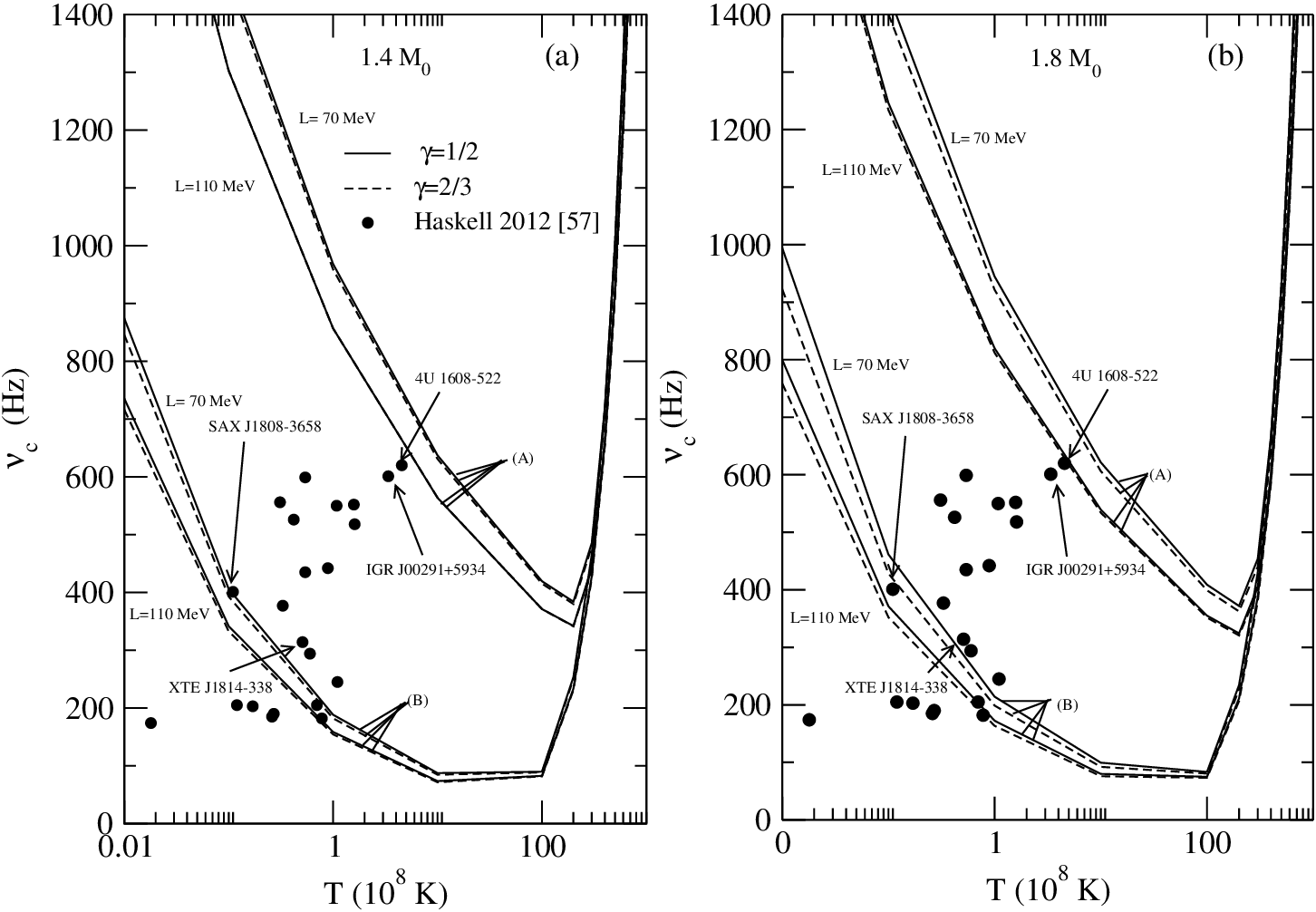}
					\end{center}
					\caption{ (a) Influence of nuclear matter incompressibility on the instability window 
on 1.4 $M_{\odot}$ NS pulsar. (b) Same as (a) but for 1.8$M_{\odot}$. Legends for the curves used in both panels are the same. 
For details, see the text. }
	\label{fig:graph_a12_23}
\end{figure}
 
 As a NS reaches the instability window the $r$-mode becomes unstable and the amplitude of the oscillation increases  
raising the temperature of the star. A newly born NS, whose temperature is $T\geq 10^{11}$K, enters in the region of instability and the star 
cools via neutrino and thermal emission. In the case of old accreting stars the torque acquired due to accretion of mass is mostly responsible 
for  the entrance in the instability region. The rise in the temperature of the star due to the unstable $r$-mode sets the viscous mechanism to act 
more effectively. The temperature of the star will increase till the $r$-mode amplitude attains a saturation value due to nonlinear effects. 
So far it is not clear which type of nonlinear mechanism is actually responsible for saturating the $r$-mode amplitude. Different considerations  
predict different ranges for the saturation value of the $r$-mode amplitude. For example, in the formulation where the crust is not considered 
and the suprathermal bulk viscosity is taken to be the nonlinear mechanism, the saturated value of $\alpha$ is found to be $\approx$1 
\cite{Alford2010,Alford2012}.
The study of the mode coupling performed in \cite{Arras2003,Bondarescu2007,Bondarescu2009} predicts a saturation value of $\alpha$ 
$\approx$  $10^{-4}$. In Ref.\cite{Mahmoodifar2013} Mahmoodifar and Strohmayer have calculated an upper limit of $\alpha$ 
$\approx$ $10^{-8}$ - $10^{-6}$ from the consideration that the $r$-mode heating provides the source of the NS luminosity in the 
accreting NS in the absence of accretion.

 As the $r$-mode amplitude $\alpha$ attains the saturation value, the NS emits gravitational waves and releases its angular 
momentum and energy and spins down to the region of stability. The spin-down rate can be calculated for a NS from equation 
(\ref{eq17}), provided the NS mass $M$, the temperature $T$, the initial angular velocity $\Omega_{in}$ and the $r$-mode amplitude 
$\alpha$ of the star are known. The spin-down rate is sensitive to the EOS through the quantity $Q$ (see Eq. (\ref{eq14})) and also 
crucially depends on the saturation value of the $r$-mode amplitude $\alpha$. In order to examine this, we have computed the spin-down 
rate from equation (\ref{eq17}) for 1.4 $M_{\odot}$ and 1.8 $M_{\odot}$ stars with $\Omega_{in}$=730 Hz using the EOSs $\gamma$=1/2 and 2/3. 
In each case, the spin-down rate is calculated as a function of time for NS masses of 1.4 $M_{\odot}$ and 1.8 $M_{\odot}$ with two values of 
$L$=70 and 110 MeV using two values of $\alpha$=$10^{-8}$ and 1. The results for $\alpha$=$10^{-8}$ and of $\alpha$=1 are shown in the 
upper panels and lower panels in Figure 7, respectively. It is found that the spin-down rate is much faster for $\alpha$=1 than for 
$\alpha=10^{-8}$. The spin-down rate increases with increasing $L$ value, with increasing mass of the pulsar NS, as well as with growing nuclear 
matter incompressibility. In order to have a quantitative idea on the spin-down rate for different $\alpha$ values, here we take the 
example of the particular pulsar NS 4U 1608-522 whose frequency is 620 Hz. According to the minimal model it is in the unstable region. 
The period needed to reach the instability boundary can be calculated from equation (\ref {eq18}) where $\Omega_{in}$ is 620 Hz. 
If the NS mass is of 1.4 $M_{\odot}$, the period to reach the boundary of the EOS, $\gamma$=1/2 and $L$=70 MeV, of the minimal model in 
figure 3(a) is 6.3$\times 10^{21}$ yr if $\alpha =10^{-8}$, whereas it will decrease to 5.7$\times 10^5$ yr if $\alpha$=1.
\begin{figure}[ht]
\vspace{1.0 cm}
        \begin{center}
        \includegraphics[width=0.9\columnwidth,angle=0,clip=true]{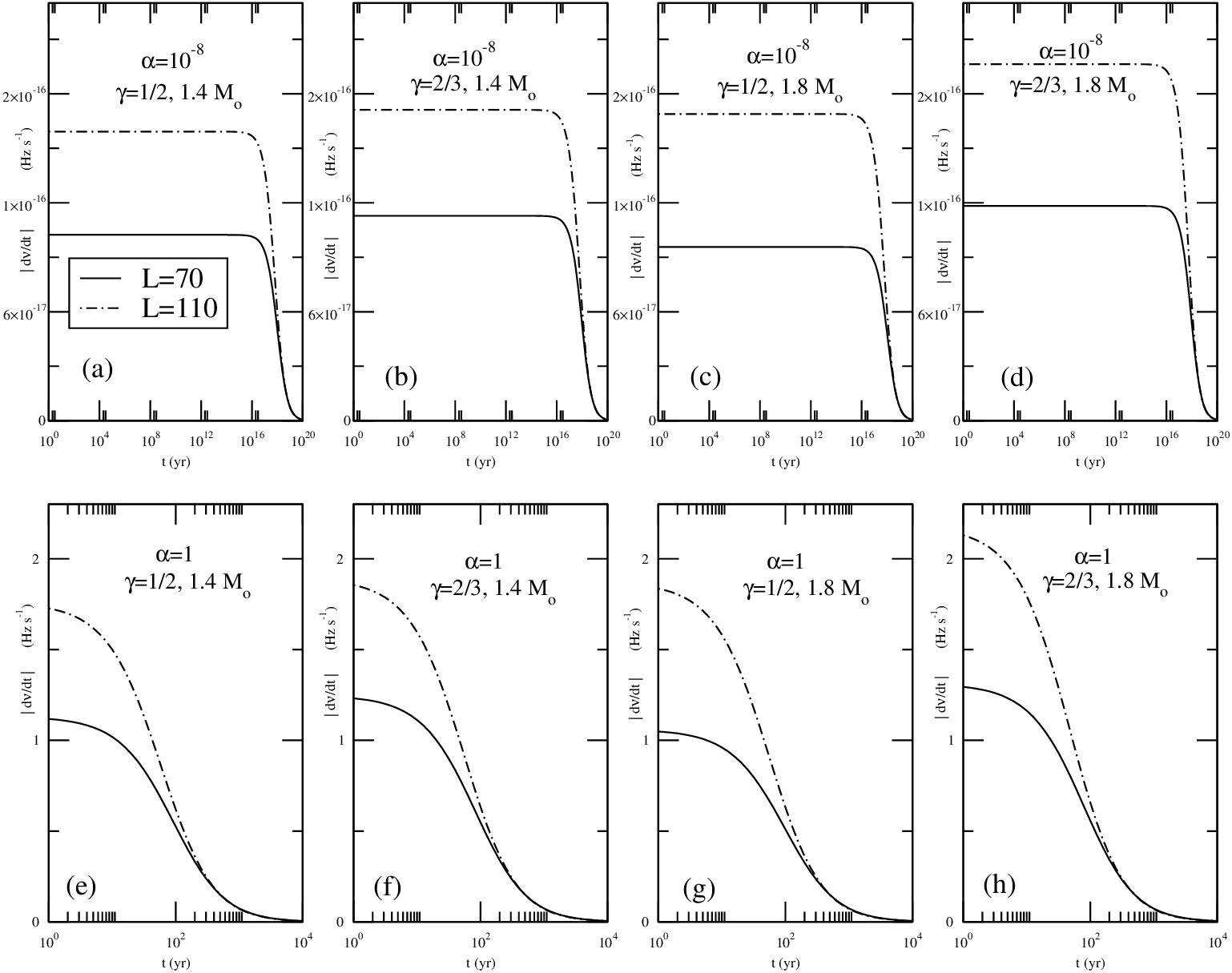}
                        \end{center}
        \caption {(upper panel) The spin-down rate as a function of time (yr) for 1.4 $M_{\odot}$ and 1.8 $M_{\odot}$ NS in case of the 
two EOSs $\gamma$= 1/2 and 2/3 with two values of $L$ =70 and 110 MeV in each case calculated from equation (\ref{eq14}) where $\alpha$= 
$10^{-8}$ is used. (lower panel)(b) Same as upper panel but for $\alpha$= 1. Legends used for the curves in all panels 
are the same. }
                \label{fig:dnudt_alpha2}
\end{figure}

We shall now calculate the limiting value of the $r$-mode amplitude $\alpha$ from the considerations of 'spin equilibrium' and 
'thermal equilibrium'. The spin equilibrium \cite{Brown2000,Ho2011,Haskell2011,Watts2008,Mahmoodifar2013} is based on the assumption that 
the outburst-quiescence cycle is balanced by the $r$-mode spin-down torque due to gravitational radiation in the whole cycle. The condition 
resulting from this is given by \cite{Mahmoodifar2013},
\begin{equation}
2\pi {\dot{\nu}_{su}} \Delta=\frac{2J_c}{\tau_{GR}},
\label{eq41}
\end{equation}
where, $J_c=-\frac{3}{2}{\Omega}{\alpha^2}{\widetilde{J}}{M}{R^2}$ is the canonical angular momentum of the $r$-mode; ${\dot{\nu}_{su}}$ 
is the spin up rate during outburst and $\Delta=t_0/t_r$ is the ratio of the outburst duration ($t_0$) to the recurrence time ($t_r$). 
Since the values of $\Delta$ and ${\dot{\nu}_{su}}$ entering in the left hand side of equation (\ref{eq41}) can be 
extracted from observations of LMXBs, these quantities can be used to compute $J_c$ in the right hand side of (\ref{eq41}).
%assuming that $\tau_{GR}$ is known. 
The $\alpha$ value which appears in $J_c$ can now be computed for a given EOS.
%From the obtained value of $J_c$ and using the expression of its definition, one can compute
%the amplitude of the $r$-mode corresponding to the spin-equilibrium condition for a given EOS. the $\alpha$ value which appears in $J_c$ can be computed for a given EOS.
%Since $\Delta$ and ${\dot{\nu}_{su}}$, entering in the left hand side of equation (\ref{eq41}), can be 
%extracted from the observational LMXBs, the $\alpha$ value which appears in $J_c$ can be computed for a given EOS.
Using the available data for $\Delta$ and $\dot{\nu}_{su}$ of the three pulsar NSs, namely IGR J00291+5934, SAXJ1808-3658 and XTE J1814-338, the 
values of the $r$-mode amplitudes $\alpha$, under the spin equilibrium consideration, obtained for EOSs $\gamma=$1/2 and 2/3 with their characteristic $L$ values 76.26 MeV and 77.31 MeV are 
given in Table 4 for NS masses of 1.4 $M_{\odot}$ and 1.8 $M_{\odot}$. The results obtained with 1.4 $M_{\odot}$ can be directly compared 
with the corresponding results in table 2 of Ref. \cite{Mahmoodifar2013} computed using the microscopic EOS of Akmal, Pandharipande and 
Ravenhall (APR) \cite{APR1998}. Our predictions for $\alpha$ in this work are in the range $10^{-7}$, which is in close agreement with the values found in 
Ref. \cite{Mahmoodifar2013}. From the results for 1.4 $M_{\odot}$ and 1.8 $M_{\odot}$ in table 4, it can be seen that $\alpha$ decreases 
with increasing NS mass as well as with increasing incompressibility of the EOS.
In order to see the influence of the $L$ value, we have calculated $\alpha$ from equation (\ref{eq41}) for $L$ = 70 MeV and 110 MeV in 
1.4 $M_{\odot}$ and 1.8 $M_{\odot}$ stars, and it is found that $\alpha$ decreases with an increase in $L$.
\begin{table}[ht]
\begin{center}
\caption{Upper bound on $r$-mode amplitude $\alpha_{sp.eq}$ from the spin equilibrium (sp.eq) condition for the EOSs $\gamma=$1/2 and 2/3 
with characteristic $L$ values 76.26 MeV and 77.31 MeV. The data for $\Delta$ and ${\dot{\nu}_{su}}$ are taken from Ref. \cite{Mahmoodifar2013}. }
\renewcommand{\tabcolsep}{0.0991cm}
\renewcommand{\arraystretch}{1.2}
\begin{tabular}{|c|c|c|c|c|c|c|}\hline
Source  														& 
$\Delta=\frac{t_0}{t_r}$            & 
${\dot{\nu}_{su}}$   & $\alpha_{sp.eq}$    & 
$\alpha_{sp.eq}$& $\alpha_{sp.eq}$  & 
$\alpha_{sp.eq}$										\\
          													&                         
          													& 
Hz $s^{-1}$ 												& 
1.4 $M_{\odot}$         						& 
1.8 $M_{\odot}$											& 
1.4 $M_{\odot}$              				& 
1.8 $M_{\odot}$											\\
                 										&                                              
                 										& 
                 										& 
for $\gamma=1/2$    								& 
for $\gamma=1/2$ 										&    
for $\gamma=2/3$        						& 
for $\gamma=2/3$										\\\hline
                                                               
IGR J00291+5934&
$\frac{13}{1363}$      &$5.0 \times10^{-13}$&$1.2980\times10^{-7}$&$1.2764\times10^{-7}$&$1.2292\times10^{-7}$ & $1.1589\times10^{-7}$    \\\hline
SAX J1808-3658&
$\frac{40}{2\times365}$& $2.5 \times10^{-14}$&$2.8622\times10^{-7}$&$2.8145\times10^{-7}$&$2.7105\times10^{-7}$ & $2.5554\times10^{-7}$ \\\hline
XTE J1814-338&
$\frac{40}{19\times365}$&$1.5 \times10^{-14}$&$1.8742 \times10^{-7}$&$1.8430\times10^{-7}$ &$1.7749\times10^{-7}$ &$1.6734\times10^{-7}$ \\\hline
\end{tabular}
\end{center}
\end{table}

We now compute the amplitude $\alpha$ from the thermal equilibrium condition. The thermal steady state during the spin down of the NS is a 
rigorous result when the mode is saturated and, in particular, it is independent of the cooling mechanism \cite{Alford2012}. 
In a steady state the gravitational radiation pumps energy into the $r$-mode at a rate given by \cite{Mahmoodifar2013}
\begin{equation}
W_d=\frac{1}{3}\Omega J_c = -2\frac{\tilde{E}}{\tau_{GR}}, 
\label{eq420}
\end{equation}
from where, by taking into account the explicit expression for $J_c$ given before, one can write the amplitude $\alpha$ in thermal
equilibrium as
\begin{equation}
\alpha=\left[\frac{-\tau_{GR} W_d} {\widetilde{J} M}\right]^{1/2}\frac{1}{\Omega R}.
\label{eq42}
\end{equation}
\begin{table}[ht]
\begin{center}
\caption{$r$-mode amplitude $\alpha_{th.eq}$ from the thermal equilibrium (th.eq) condition for the EOSs, $\gamma=$1/2 with $L$ values 70 MeV and 110 MeV.}
\renewcommand{\tabcolsep}{0.0991cm}
\renewcommand{\arraystretch}{1.2}
\begin{tabular}{|c|c|c|c|c|c|}\hline
Source	  & $\alpha_{th.eq}$  & $\alpha_{th.eq}$ & $\dot{\nu}$ Hz $s^{-1}$&$\dot{\nu}$ Hz $s^{-1}$&$\dot{\nu}$ Hz $s^{-1}$\\
		      & 1.4 $M_{\odot}$		&  1.8 $M_{\odot}$ & 1.4 $M_{\odot}$		&  1.8 $M_{\odot}$          & Observation\\\hline
\multicolumn{5}{|c|}{$\gamma=1/2$, $L=70$ MeV}&\multicolumn{1}{|c|}{}   \\\hline 
4U1608-522          &7.3144$\times10^{-8}$ &6.6442 $\times10^{-8}$& -1.0969$\times10^{-14} $ & -8.4609$\times10^{-15}$&\\\hline
IGR J00291+5934     &1.4644$\times10^{-8}$ &1.3302 $\times10^{-8}$& -3.4542 $\times10^{-16}$ & -2.6645$\times10^{-16}$&$-3\times10^{-15}$   \\\hline
MXB 1659-29         &1.1838$\times10^{-8}$ &1.0753 $\times10^{-8}$& -1.3401 $\times10^{-16}$ & -1.0336$\times10^{-16}$&\\\hline
Aql X-1             &3.6112$\times10^{-8}$ &3.2803 $\times10^{-8}$& -1.1558 $\times10^{-15}$ & -8.9158$\times10^{-16}$&\\\hline
KS 1731-260         &2.4306$\times10^{-8}$ &2.2079 $\times10^{-8}$& -3.7308 $\times10^{-16}$ & -2.8778$\times10^{-16}$&\\\hline
XTE J1751-305				&5.2651$\times10^{-8}$ &4.7827 $\times10^{-8}$& -4.7565 $\times10^{-16}$ & -3.6690$\times10^{-16}$ &$-5.5\times10^{-15}$ \\\hline
SAX J1808-3658			&1.3017$\times10^{-8}$ &1.1824 $\times10^{-8}$& -1.6447 $\times10^{-17}$ & -1.2686$\times10^{-17}$ &$-5.5\times10^{-16}$\\\hline
XTE J1814-338				&1.8316$\times10^{-7}$ &1.6638 $\times10^{-7}$& -5.8778 $\times10^{-16}$ & -4.5339$\times10^{-16}$&	 \\\hline
NGC 6440						&1.6028$\times10^{-6}$ &1.4559 $\times10^{-6}$& -2.2755 $\times10^{-15}$ & -1.7552$\times10^{-15}$&\\\hline
\multicolumn{5}{|c|}{$\gamma=1/2$, $L=110$ MeV}& \multicolumn{1}{|c|}{} \\\hline 
4U1608-522          &5.8984$\times10^{-8}$ &5.0093 $\times10^{-8}$& -1.1091 $\times10^{-14}$ & -8.4892$\times10^{-15}$&\\\hline
IGR J00291+5934     &1.1809$\times10^{-8}$ &1.0029 $\times10^{-8}$& -3.4930  $\times10^{-16}$ & -2.6734$\times10^{-16}$&$-3\times10^{-15}$   \\\hline
MXB 1659-29         &9.5462$\times10^{-9}$ &8.1072 $\times10^{-9}$& -1.3551 $\times10^{-16}$ & -1.0371$\times10^{-16}$&\\\hline
Aql X-1             &2.9121$\times10^{-8}$ &2.4731 $\times10^{-8}$& -1.1688 $\times10^{-15}$ & -8.9457$\times10^{-16}$&\\\hline
KS 1731-260         &1.9600$\times10^{-8}$ &1.6646 $\times10^{-8}$& -3.7727 $\times10^{-16}$ & -2.8875$\times10^{-16}$&\\\hline
XTE J1751-305				&4.2458$\times10^{-8}$ &3.6058 $\times10^{-8}$& -4.8099 $\times10^{-16}$ & -3.6814$\times10^{-16}$ &$-5.5\times10^{-15}$ \\\hline
SAX J1808-3658			&1.0497$\times10^{-8}$ &8.9147 $\times10^{-9}$& -1.6631 $\times10^{-17}$ & -1.2729$\times10^{-17}$ &$-5.5\times10^{-16}$\\\hline
XTE J1814-338				&1.4770$\times10^{-7}$ &1.2543 $\times10^{-7}$& -5.9437 $\times10^{-16}$ & -4.5491$\times10^{-16}$&	 \\\hline
NGC 6440						&1.2924$\times10^{-6}$ &1.0976 $\times10^{-6}$& -2.3010 $\times10^{-15}$ & -1.7611$\times10^{-15}$&\\\hline
\end{tabular}
\end{center}
\end{table}
Hence, in the thermal steady state all the energy emitted from the star during the quiescence is due to the $r$-mode dissipation 
inside the star. The thermal equilibrium condition implies that $W_d=L_{\nu}+L_{\gamma}$, where $L_{\nu}$ and $L_{\gamma}$ are, respectively, the neutrino 
luminosity and the thermal photon luminosity at the surface of the star. Assuming standard neutrino cooling, the thermal
equilibrium condition for a NS can be approximated by $W_d \simeq L_{\gamma}$, since the luminosity due to the neutrino cooling 
can be neglected compared to the surface photon luminosity in not too heavy NSs ($M<2 M_{\odot}$) \cite{Mahmoodifar2013}.
Thus the amplitude $\alpha $ is computed using $L_{\gamma}=4 \pi R^{2} \sigma T_{eff}^{4}$, where $\sigma$ is the Stefan's 
constant and $T_{eff}$ is the effective surface temperature of the star. Under this approximation, we obtain
\begin{equation}
\alpha=7.9494\times 10^{-17} \left[\frac{-\tau_{GR} } {\widetilde{J}}\right]^{1/2}   \frac{\sigma^{1/2} T_{eff}^{2}}{\Omega} 
\left[\frac{M_{\odot}}{M}\right]^{1/2},
\label{eq126}
\end{equation}
for the $r$-mode amplitude in the case of the thermal equilibrium consideration.

For the EOS $\gamma=$1/2 and $L$ in the range 70-110 MeV, $ \alpha$ is calculated from equation (\ref{eq126}) for the pulsar NSs, which are 
predicted to lie in the unstable region in figure 3, using the data for $T_{eff}$ taken from table 2 of Refs. \cite{Heinke2007,Heinke2009}. The 
results are given in Table 5. The predicted $\alpha$ values derived in this work from the consideration of thermal equilibrium, lie in 
the range of $10^{-8}$ - $10^{-7}$, which is in agreement with the results obtained in table 4 of Ref.\cite{Mahmoodifar2013}. The spin-down rates 
of these stars are now calculated from equation (\ref{eq14}), which 
depends on the value of $Q$, the mass and radius of the NS and the $r$-mode amplitude $\alpha$. The $Q$-value is not sensitive to the slope 
of the symmetry energy  $L$. For example, using the EOS with $\gamma=$1/2 the value of $Q$ varies from  0.09411 to 0.09341 (0.09670 to 0.09602)
 for a NS with mass 1.4 $M_{\odot}$ (1.8 $M_{\odot}$) as $L$ changes from 70MeV to 110 MeV.
 %For example using the EOS with $\gamma=$2/3 the value of $Q$ varies from  0.09449 to 0.093781 (0.09741 to 0.09673) for a NS with mass 
%1.4 $M_{\odot}$ (1.8 $M_{\odot}$) as $L$ changes from 70MeV to 110 MeV. 
 The $L$ dependence of $Q$ shows a similar behaviour in the case of the EOS with $\gamma=$2/3.  
Thus, $\dot{\nu}$ is sensitive to the mass and radius of the NS and $r$-mode amplitude $\alpha$. The results of $\dot{\nu}$ calculated from equation (\ref{eq14}) using the respective $\alpha$ obtained from thermal equilibrium condition for 1.4 $M_{\odot}$ and 1.8 $M_{\odot}$ mass NSs and their predicted radii are also given in table 5. The spin down rate obtained in the present case for 1.4 $M_{\odot}$ mass NSs is closer to the data for the three NS pulsars, IGR J00291+5934, XTE J1751-305 and SAX J1808-3658, in comparison to the corresponding results obtained in Ref.\cite{Mahmoodifar2013} for APR EOS. From Table 5 it can be seen that $\alpha$ decreases with an increase in $L$ as well as with an increase in the mass of 
the NS. The $\dot{\nu}$ depends on $L$ in a similar manner as in $\alpha$. The $L$ dependence 
of $\dot{\nu}$ in the case of the EOS $\gamma=$ 2/3, for both 1.4 $M_{\odot}$ and 1.8 $M_{\odot}$, has been verified to be similar to the one 
obtained for $\gamma=$1/2. The $\dot{\nu}$ values for $\gamma=$2/3 are found to be almost the same to their $\gamma=$1/2 counterparts given
 in table 5. Since in all of the three NS pulsars, whose spin-down rates are measured, the predicted results lie below the measured values, 
the mass measurement of these NSs become essential before going for exploring other possible modifications.

\section{Summary and conclusions}
\label{Sec:con}
 The characteristic features of the $r$-mode oscillation in rotating neutron stars are outlined. The EOS of pulsar NSs is constructed under 
the consideration of a core composition of normal neutron, proton, electron and muon matter in charge neutral $\beta$-equilibrium condition. The 
theoretical EOS based on the finite range simple effective interaction is used to this end. The $r$-mode instability window for the pulsar NS 
is computed by taking into account dissipation by shear and bulk viscosities and by the viscous layer at the crust-core 
boundary. The influence of the 
slope parameter of the symmetry energy $L(\rho_0)$ and the  nuclear matter incompressibility $K(\rho_0)$ on the $r$-mode window have been 
studied. In the rigid crust-core model ($S$=1) the viscous layer dissipation at the crust-core boundary is the dominant mechanism to decide 
the critical frequency in the temperature range $T\leq$ $10^{10}$ K. In this range of $T$ the instability window lowers for higher $L$ values. 
This has been illustrated in figure 4 as due to the effect of $L$ on the time-scales associated to the viscous dissipation at the crust-core 
boundary layer and to the gravitational radiation, $\tau_{VE}$ and $\tau_{GR}$, respectively.
In the temperature range $T>$ $10^{10}$ K, the time-scale associated to the bulk 
viscosity ($\tau_{BV}$) is dominant and remains almost insensitive to the $L$-value. Therefore the 
region of instability increases with increasing $L$, primarily due to its influence on $\tau_{VE}$ and $\tau_{GR}$. Under the consideration
 of a rigid-crust core, none of the neutron stars of LMXBs and MSRPs displayed in Figure 3 are predicted to be unstable. However, if the 
penetration of the core into the crust is taken into account in terms of a slippage factor with realistic value, $S$=0.05, many NS pulsars 
(14 shown in figure 3) are predicted to be unstable. 
This result is quite a similar to the one
found in Ref \cite{Haskell2012}. The upper bound of the $r$-mode amplitude is estimated from the spin-equlibrioum condition for acreting NSs based on the assumption that the acreting torque is balanced by the $r$-mode gravitational torque.
Using the data of three pulsar NSs, the upper bound of the of the $r$-mode amplitude $\alpha$ is found to be of the order $10^{-7}$ for the NSs masses 1.4 $M_{\odot}$ to 1.8 $M_{\odot}$. However the quiscent luminosity in some of the NSs contradict these results. The calculated luminosity with the $\alpha$ obtained from spin equilibrium condition predicts larger value. 
The upper bound on $\alpha$ is therefore calculated from the thermal equilibrium condition. Since spin down occurs in thermal steady state, the heat generated by the $r$-mode processes equals to the energy radiated from the NS. Under the approximation that the energy radiated by modified URCA process in not too heavy NS is small in comparison to the thermal radiation from the surface of the NS, the upper bound of $\alpha$ is found in the range $10^{-8}$ to$10^{-6}$ for the stars in the unstable region. The spin-down rate of these stars are also calculated. 
Our predictions in case of three pulsar stars, for which data on spin-down rate exist, are closer to the data as compared to the similar results reported in earlier literature. The $\alpha$ value decreases when the slope of the symmetry 
energy increases as well as for larger NS mass. The spin-down rate also crucially depends on the mass and radius of the NS pulsar, apart from the $\alpha$ value. All these facts warrant the mass measurement of the NSs. 
 
 In the study of thermal equilibrium, only the conventional thermoluminosity process is considered as the mechanism for radiation of heat 
energy produced inside the star. It is necessary to include the energy radiation by the modified URCA cooling and also by the direct URCA 
cooling, wherever the later process is physically allowed. The effects due to the superfluidity of neutrons and superconducting protons 
also need to be taken into account. Without these considerations the results on $\alpha$ and $\dot{\nu}$ obtained here are only qualitative. 
The influence of the presence of hyperons and a possible phase transition to quark matter in the NS core also need to be examined. 
Uncertainties on these aspects and possible processes that might be taking place are mainly concerned with our poor knowledge on the 
composition of the core of NSs. The mass measurements of LMXB NSs together with the observational data can be of paramount importance in the 
direction of resolving the uncertainty on the core composition. This will, in turn, also help in answering at least some of the existing 
queries on the EOS of highly neutron-rich dense nuclear matter.

\section*{Acknowledgments}
One of the authors, TRR, acknowledges the Institute of Cosmos Sciences of the University of Barcelona (ICCUB) for the support and hospitality 
during his visit. M.C. and X.V. acknowledge support from Grant FIS2014-54672-P from MINECO and FEDER, Grant 2014SGR-401 from Generalitat 
de Catalunya and the project MDM-2014-0369 of ICCUB (Unidad de Excelencia Maria de Maeztu) from MINECO.

\section*{References}

\end{document}